\documentclass[journal]{IEEEtran}
\usepackage{pgfplotstable}

\usepackage[english]{babel}
\usepackage[utf8x]{inputenc}
\usepackage[T1]{fontenc}

\usepackage{amsmath,amssymb,mathrsfs,bbm}
\usepackage{graphicx}
\usepackage[colorinlistoftodos]{todonotes}
\usepackage[colorlinks=true, allcolors=blue]{hyperref}
\usepackage{multirow}
\usepackage{makecell}
\usepackage[lined,linesnumbered,ruled]{algorithm2e}

\usepackage{psfrag,epsfig,graphics}
\usepackage{amsmath,amsthm,amssymb,multirow}
\usepackage{mathbbol}
\usepackage{amssymb}   
\usepackage{url}  
\usepackage{xstring}
\usepackage{cite}

\usepackage{mathabx} %

\DeclareSymbolFontAlphabet{\amsmathbb}{AMSb}%

\newcommand{\cp}[1]{\ifmmode {\mathcal{#1}}\else ${\mathcal{#1}}$\fi}
	
\newcommand{\bA}{\boldsymbol{A}}

\newcommand{\bI}{\boldsymbol{I}}

\newcommand{\bM}{\boldsymbol{M}}

\newcommand{\bW}{\boldsymbol{W}}
\newcommand{\bX}{\boldsymbol{X}}
\newcommand{\bY}{\boldsymbol{Y}}
\newcommand{\bZ}{\boldsymbol{Z}}

\newcommand{\ba}{\boldsymbol{a}}
\newcommand{\bb}{\boldsymbol{b}}

\newcommand{\bm}{\boldsymbol{m}}

\newcommand{\bh}{\boldsymbol{h}}

\newcommand{\be}{\boldsymbol{e}}

\newcommand{\by}{\boldsymbol{y}}

\newcommand{\bx}{\boldsymbol{x}}

\newcommand{\bz}{\boldsymbol{z}}

\newcommand{\diag}{\operatorname{diag}}

\newcommand{\calD}{\mathcal{D}}

\newcommand{\calL}{\mathcal{L}}
\newcommand{\calN}{\mathcal{N}}
\newcommand{\calR}{\mathcal{R}}

\newcommand{\bepsilon}{\boldsymbol{\epsilon}}
\newcommand{\bgamma}{\boldsymbol{\gamma}}

\newcommand{\bmu}{\boldsymbol{\mu}}
\newcommand{\bsigma}{\boldsymbol{\sigma}}

\newcommand{\bxi}{\boldsymbol{\xi}}

\newcommand{\bSigma}{\boldsymbol{\Sigma}}

\newcommand{\bXi}{\boldsymbol{\Xi}}

\newcommand{\cb}[1]{\boldsymbol{#1}}

\newcommand\tensor[1]{%
  \ifcat\noexpand#1\relax %
    \mathbb{#1}%
  \else
      \if\relax\detokenize\expandafter{\romannumeral-0#1}\relax  %
        \mathbb{#1}
      \else
        \mathcal{#1}%
      \fi
  \fi }

\newcommand{\Ex}{\amsmathbb{E}}

\newcommand{\Dir}{\operatorname{Dir}}

\usepackage{color}  %

\definecolor{darkgreen}{rgb}{0., 0.4, 0.}

\usepackage{booktabs}

\usepackage{tikz}
\usetikzlibrary{bayesnet}
\usetikzlibrary{arrows}
\usetikzlibrary{positioning}

\title{Learning Interpretable Deep Disentangled Neural Networks for Hyperspectral Unmixing}

\author{Ricardo Augusto Borsoi, Deniz Erdo{\u{g}}mu{\c{s}, Tales Imbiriba}
\thanks{This work was supported in part by the National Geographic Society under Grant NGS-86713T-21.}
\thanks{A preliminary version of this work was presented in~\cite{borsoi2023deepDisentangledUnmixingICASSP}. This extended version contains derivations for a tractable solution to the optimization problems, interpretable network architectures, a self-supervised learning scheme, and an extended experimental validation.}
\thanks{R.A. Borsoi is with the Universit\'e de Lorraine, CNRS, CRAN, Vandoeuvre-l\`es-Nancy, F-54500, France. e-mail: \{raborsoi\}@gmail.com.}
\thanks{D. Erdo{\u{g}}mu{\c{s}} and T. Imbiriba are with the ECE department of the Northeastern University, Boston, MA, USA. e-mail: \{erdogmus, talesim\}@ece.neu.edu.}}

\begin{document}

\maketitle

\begin{abstract}
Although considerable effort has been dedicated to improving the solution to the hyperspectral unmixing problem, non-idealities such as complex radiation scattering and endmember variability negatively impact the performance of most existing algorithms and can be very challenging to address. Recently, deep learning-based frameworks have been explored for hyperspectral umixing due to their flexibility and powerful representation capabilities. However, such techniques either do not address the non-idealities of the unmixing problem, or rely on black-box models which are not interpretable. In this paper, we propose a new interpretable deep learning method for hyperspectral unmixing that accounts for nonlinearity and endmember variability. 
The proposed method leverages a probabilistic variational deep-learning framework, where disentanglement learning is employed to properly separate the abundances and endmembers. The model is learned end-to-end using stochastic backpropagation, and trained using a self-supervised strategy which leverages benefits from semi-supervised learning techniques. Furthermore, the model is carefully designed to provide a high degree of interpretability. This includes modeling the abundances as a Dirichlet distribution, the endmembers using low-dimensional deep latent variable representations, and using two-stream neural networks composed of additive piecewise-linear/nonlinear components.
Experimental results on synthetic and real datasets illustrate the performance of the proposed method compared to state-of-the-art algorithms.
\end{abstract}

\begin{IEEEkeywords}
Hyperspectral data, spectral unmixing, neural networks, disentanglement, deep learning.
\end{IEEEkeywords}

\section{Introduction}

\subsection{Background}
Due to physical limitations of the imaging process, hyperspectral images (HIs) provide very high spectral resolution but low spatial resolution, which means that a pixel usually contains a mixture of several different materials~\cite{Bioucas-Dias-2013-ID307}. %
Hyperspectral unmixing (HU) consists in estimating the spectral signatures of pure materials (i.e., \emph{endmembers} -- EMs) in a scene and the proportions with which they are contained in each pixel (i.e., \emph{abundances}) directly from an HI~\cite{Keshava:2002p5667}. Due to the unsupervised nature of HU, adequately exploring the physics of the problem when devising modeling strategies is paramount for obtaining stable and high-quality EM and abundance estimations.
Simplistic methods considered the interaction between light and the EMs to be linear~\cite{Keshava:2002p5667}. However, this model is over-simplified, degrading the quality of the estimates. Thus, addressing important non-idealities such as nonlinear interactions between light and the materials~\cite{Dobigeon-2014-ID322} and the variability of the EMs in different HI pixels~\cite{borsoi2020variabilityReview} has become the subject of much attention more recently.

Different approaches have been proposed to perform HU (see Section~\ref{sec:relatedw}). However, traditional methods often lack the flexibility to represent non-idealities observed in practical HIs. This motivated the use of machine learning approaches for HU that ally both flexibility and performance~\cite{bhatt2020deepLearningHUreview,palsson2022unmixingAECcomparison}. 
Nonetheless, interpretability remains a key point when leveraging machine learning strategies in HI analysis~\cite{hong2021interpretableAI_hyperspectral}.

Recently, physically motivated machine learning approaches have been successfully applied to HU~\cite{Chen-2013-ID321,hong2021egu_net,borsoi2019deepGun,qian2020modelInspiredNNs_SU}. The advantage of such models with respect to fully black-box strategies lies in the interpretability of the estimated EMs and abundance parameters, which is a requirement for meaningful unmixing results as it allows a user to understand the causes for the algorithm's behavior~\cite{miller2019explanationAI}. 
When deep learning strategies come into play, autoencoder (AEC) architectures are of special interest due to the intrinsic low-dimensionality of the abundance space with respect to the pixels, and to the connection between such strategies and hyperspectral mixing models~\cite{guo2015autoencodersUnmixing,palsson2018autoencoderUnmixing_IEEEaccess}. Thus, several approaches using AECs were proposed to solve HU addressing phenomena such as nonlinearity~\cite{hong2021egu_net,wang2019AECnlin,li2021modelBasedAECsSU}, EM variability~\cite{borsoi2019deepGun,shi2021generativeModelEMvariability} and outliers~\cite{su2019deepAutoencoderUnmixing}.

Although such deep learning methods presented relevant solutions for HU reaching high levels of accuracy while retaining physical interpretation, such strategies fail to provide a separation between EMs and abundances that is both interpretable and accounts for existing spectral variability and nonlinear effects.
Recently, supervised disentanglement learning has become a popular approach to separate latent variables in deep learning models into different factors of variation that can have a physical interpretation~\cite{siddharth2017disentangledVAEs_NIPS}. Disentangled decompositions have been considered for different applications (e.g., separating content from style in images~\cite{siddharth2017disentangledVAEs_NIPS}), and its potential will be explored in this work to aid the separation between abundance and EM variations in HU.

\subsection{Contributions of this work}
In this work, we aim to develop a learning-based unmixing algorithm which accounts for challenges including both nonlinearities in the mixing process and spectral variability of the EMs. 
Moreover, different from black-box models, the motivation of this work is to develop a framework that is interpretable and based on clearly defined statistical hypotheses and training criterion. Thus, we propose a self- and semi-supervised learning-based approach for HU which achieves disentanglement through independence assumptions.

In the inference strategy, the model parameters are learned by maximizing a lower bound to the likelihood of both training and test data, leading to a semi-supervised strategy. To deal with the intractable nature of the posterior distribution of the abundances and EMs given the observed pixels, we consider a variational deep learning-based strategy coupled with importance sampling to obtain a tractable objective that can be optimized efficiently using stochastic backpropagation. The proposed method is named ID-Net (\emph{Interpretable Disentangled neural Networks for HU}).
    
To model EM variability we consider a model consisting of a learnable nonlinear transformation of a low-dimensional Gaussian random variable, which can capture the low-dimensionality of the EM variability manifold~\cite{borsoi2019deepGun,borsoi2019EMlibManInterpVAE}.
We also consider the Dirichlet distribution to model the abundances in order to account for physical constraints. To tackle this distribution in a variational deep learning framework, instead of using inefficient approximations such as rejection samplers~\cite{naesseth2017rejectionSamplVAE} we consider a pathwise derivative estimator, which provides unbiased gradient estimates~\cite{jankowiak2018pathwiseDerivatives}. Furthermore, to promote sparsity of the abundances we penalize the Dirichlet's concentration parameters with an L$_{1/2}$-norm based regularization~\cite{qian2011unmixing_L12_NMF}.
To represent nonlinear mixing effects, we use a mixing model comprised of the combination of the linear mixing model with a (learnable) additive nonlinear term, which allows us to control the amount of nonlinearity in the model by means of regularization techniques, while maintaining a nonparametric representation.

Given the model, unmixing is performed using a deep variational inference framework, which computes the parameters of the model and an approximate posterior distribution. This is achieved by maximizing a lower bound on the likelihood of training and test data in a semi-supervised setting obtained using importance sampling~\cite{kingma2014semiSupVAEs,siddharth2017disentangledVAEs_NIPS}. Optimization is performed using stochastic backpropagation methods such as Adam~\cite{kingma2014adam}.
Disentanglement learning is employed to simplify the model by incorporating principled statistical independence assumptions in the variational posterior distribution, which represents the inference (unmixing) process. This allows us to remove the influence of the independent variables when parametrizing each of the posterior PDFs.
To address the scarcity of training data with ground truth for HU, we consider a self-supervised approach in which the supervised training data is generated directly from the observed HI based on image-based spectral library construction approaches~\cite{borsoi2020variabilityReview}.

Different from black box strategies our approach leads to a flexible yet interpretable nonlinear model capable of incorporating EM variability, where non-ideal effects (variability and nonlinearity) can be controlled through appropriately designed regularizations over abundance and EM estimates. Codes for the proposed method are available at \url{https://github.com/ricardoborsoi/IDNet_release}.

\begin{figure*}
    \centering
    \includegraphics[width=1\linewidth]{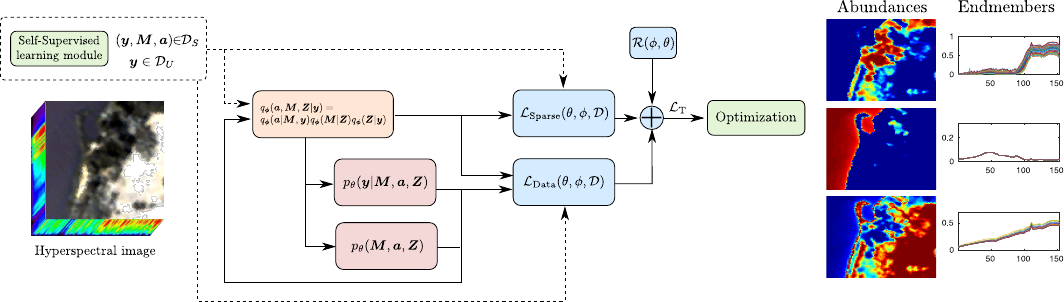}
    \vspace{-4ex}
    \caption{Overview of the proposed method. A self-supervised learning module is used to generate supervised and unsupervised training data, $\calD_U$ and $\calD_S$, to train the model. A generative model (with parameters $\theta$) is constructed with a prior model $p_{\theta}(\bM,\ba,\bZ)$, describing the a priori behavior of the EMs, abundances, and variability parameters, and a mixing model $p_{\theta}(\by|\bM,\ba,\bZ)$, containing both a linear and a nonlinear component. To solve HU, an approximate posterior distribution (with parameters $\phi$) $q_{\phi}(\bM,\ba,\bZ|\by)$ is constructed, and independence assumptions between the abundances and the EM variability parameters are used to disentangle these variables. A semi-supervised learning-based optimization objective is used to learn the parameters of both the generative model and of the approximate posterior by maximizing a lower bound on the log-likelihood of both the supervised and unsupervised data, in $\calL_{\rm Data}(\theta,\phi,\calD)$, and also regularizations promoting abundance sparsity ($\calL_{\rm Sparse}(\theta,\phi,\calD)$) and controlling the amount of nonlinearity in both the generative and inference models. The learned posterior then provides the HU solution as the EMs and the abundances.}
    \label{fig:overview}
\end{figure*}

\section{Related work}
\label{sec:relatedw}

Given an HI with $L$ bands and $N_U$ pixels $\by_n\in\amsmathbb{R}^L$, HU aims to recover the signatures of the $P$ EMs in the scene $\bM\in\amsmathbb{R}^{L\times P}$ and their abundances $\ba_n\in\amsmathbb{R}^P$ for each pixel $n\in\{1,\ldots,N_U\}$.
The linear mixing model (LMM) is commonly used to represent the interaction between light and the materials in the scene~\cite{Keshava:2002p5667}. However, several non-idealities are observed in practice, such as nonlinear interactions between light and the materials~\cite{Dobigeon-2014-ID322} and the variability of the EM spectra for each HI pixel~\cite{borsoi2020variabilityReview}. This requires more general models, which can represent the mixing process as:
\begin{align}
    \label{eq:general_mm}
    \by_n = \boldsymbol{f}(\ba_n,\bM,\bxi_n) + \be_n \,,
\end{align}
where function $\boldsymbol{f}$ describes the mixing process, $\bxi_n$ is a vector of parameters which account for nonlinearity and spectral variability, and $\be_n\in\amsmathbb{R}^L$ denotes additive noise. 
Different unmixing approaches have been proposed using the LMM. These strategies were based on frameworks including, e.g., nonnegative matrix factorization~\cite{qian2011unmixing_L12_NMF,yao2019nonconvexSparsityNonlocalSmoothUnmixing}, Bayesian estimation~\cite{dobigeon2008}, sparse regression~\cite{iordache2011sunsal,borsoi2018superpixels1_sparseU} and tensor decomposition~\cite{qian2016NTF_HU,imbiriba2018ULTRA_V,borsoi2019tensorInterpolationICASSP}.
However, more recent methods have been focused on addressing non-idealities in the model and on leveraging recent advances in machine learning.

\paragraph*{\textbf{Nonlinear HU}}

Nonlinear HU accounts for complex radiation scattering among several EMs which occurs in many scenes (e.g., vegetated areas~\cite{Ray1996}).
Such strategies can be generally divided between model-based and model-free approaches. Model-based approaches rely on prior knowledge about the mixing process, including the bilinear mixing model (BLMM)~\cite{Nascimento2009,halimi2011}, the post-nonlinear mixing model (PNMM)~\cite{Altmann_tip_2012}, and Hapke's model~\cite{HapkeBook1993}. 
On the other hand, model-free methods attempt to learn the nonlinearity directly from the observed data using, e.g., support vector machines~\cite{Li_svm_2005}, algorithms based on graph geodesic distances~\cite{Heylen:2011kc} and kernel-based methods~\cite{Chen-2013-ID321,borsoi2020BMUAN}. Recent approaches have relied on deep learning techniques, reviewed in the following.

\paragraph*{\textbf{HU with endmember variability}}

The variability of the EMs in a scene is one of the main challenges in HU~\cite{borsoi2020variabilityReview}. Existing solutions represent the EMs either as sets of spectral signatures, as statistical distributions, or using parametric representations of spectral variability~\cite{Zare-2014-ID324-variabilityReview}. Representing EMs as sets leads to HU being formulated as a structured sparse regression problem with convex~\cite{iordache2011sunsal} and non-convex~\cite{drumetz2019SU_bundlesGroupSparsityMixedNorms} regularization penalties, or combinatorial optimization problems~\cite{roberts1998originalMESMA,borsoi2021MT_MESMA}. Although this approach is popular in practice due to its simplicity and interpretability, it requires accurate spectral libraries to obtain a good performance.
Statistical distributions such as a Gaussian~\cite{halimi2015unsupervisedBayesianUnmixing}, Beta~\cite{du2014spatialBetaCompositional}, or a mixture of Gaussians~\cite{zhou2018variabilityGaussianMixtureModel} are also used to represent the EMs. However, realistic distributions make the HU problem hard to solve and computationally intensive.
Tensor-based approaches based on low-rank regularization~\cite{imbiriba2018ULTRA_V} or on convolutional sparse coding regularization~\cite{yao2021sparsityConvolutionalDecomposition} has also been proposed for HU accounting for the EM variability.
More recently, parametric models have become popular in HU. Such methods represent EM variations in each HI pixel using, e.g., additive perturbations~\cite{Thouvenin_IEEE_TSP_2016_PLMM}, spectrally uniform or bandwise multiplicative scalings~\cite{drumetz2016blindUnmixingELMM,imbiriba2018glmm}, combinations of additive and multiplicative factors~\cite{hong2019augmentedLMMvariability}, or multiscale spatial models~\cite{Borsoi_multiscaleVar_2018}.
HU is then formulated as a non-convex optimization problem using carefully designed regularizations to constrain the degrees of freedom of the model. Recent work also considered a linear-quadratic nonlinear mixture model with EM variability in HU~\cite{deville2023unmixing_NMF_nonlienar_variability}.

\paragraph*{\textbf{Deep learning-based HU}}

Deep learning has become an established tool for hyperspectral imaging, with methods such as graph convolutional neural networks 
(NNs)~\cite{hong2020graphNN_CGN_hyperspectralClassification} and the transformer architecture~\cite{hong2021spectralformerHyperspectralClassification} bringing significant advances to hyperspectral classification.
Various deep learning-based approaches have also been proposed for HU. %
Early works trained NNs as supervised regression methods to learn a mapping from the mixed pixels to their abundance values~\cite{guilfoyle2001comparativeUnmixingNeuralNetworksRBF,plaza2010selectingTrainingSamplesNNunmixing}. However, those methods are hindered by the limited availability of training data with ground truth. %
This motivated the development of algorithms such as extended support vector machines (i.e., soft classification)~\cite{li2015unmixingExtendedSVMgeometricAnalysis}, which require only EM signatures as training samples.

Recently, autoencoder (AEC) networks have become widely used in unsupervised HU~\cite{palsson2022unmixingAECcomparison}. AECs are encoder-decoder (i.e., bottleneck) NNs which can be applied for HU by identifying the decoder with the mixing model~\eqref{eq:general_mm}, the encoder with its inverse, and the latent codes with the abundances~\cite{guo2015autoencodersUnmixing,palsson2018autoencoderUnmixing_IEEEaccess}. AECs have been developed for linear HU using, e.g., appropriately designed encoder networks or preprocessing approaches to reduce the effect of noise and outliers~\cite{su2019deepAutoencoderUnmixing,sahoo2022HU_geological_latentEncoding}, sparsity constraints~\cite{ozkan2018endnet_autoencoderUnmixing,qu2018udas_autoencoderUnmixing}, or convolutional architectures to exploit spatial information about the HIs~\cite{palsson2020convolutionalAEC_SU,hua2021gatedAEC_SU}. %
Coupled convolutional AECs were also used to jointly unmix hyperspectral and multispectral data for image fusion~\cite{yao2020crossAttentionUnmixingNetsSuperResoluton}.

Other methods based on AECs proposed to address nonlinear HU by using nonlinear architectures for the decoder, including a post-nonlinear model~\cite{wang2019AECnlin}, additive nonlinearities~\cite{zhao2021AECnonlinearSUattitive3D,zhao2021LSTM_AEC_SU}, or using specifically designed nonlinear NN layers~\cite{shahid2021unsupervisedSUautoencoder}.
A model-based architecture for the encoder was also proposed in~\cite{li2021modelBasedAECsSU} by exploring the relationship between the AEC and the nonlinear mixing model.

AECs have also been used to address EM variability in HU as generative EM models to capture the low-dimensional manifold of spectral variability~\cite{borsoi2019deepGun}. This model was used in matrix factorization~\cite{borsoi2019deepGun}, structured sparse regression~\cite{borsoi2019EMlibManInterpVAE}, and probabilistic~\cite{shi2021generativeModelEMvariability} HU methods. Endmember variability was also addressed in~\cite{zhao2022AEC_SU_variability3D} a spatial-spectral autoencoder using an additive perturbation model~\cite{Thouvenin_IEEE_TSP_2016_PLMM}, and in non-AEC-based approaches using Gaussian Process regression~\cite{uezato2016unmixingGaussianProcessVariability,koirala2020geodesicSupervisedSUvariability}.
Sets of multiple pure pixels extracted from the HI (which can be seen as samples from EM signatures) have also been explored to improve the robustness of AECs in HU, either by learning mappings from the pure pixels to their corresponding abundances~\cite{jin2021two_stream_AEC_SU,hong2021egu_net}, or by regularizing the EMs estimated by the decoder~\cite{li2021selfSuperv_deepNMF_SU}. Recurrent NNs have also been recently employed in~\cite{borsoi2023dynamicalUnmixingVRNNs} to perform HU accounting for the temporal variability of the EMs.

Other works have also considered different loss functions for AEC training, such as the Wasserstein distance~\cite{min2021metricLearningNet_SU} and adversarial losses~\cite{jin2021adversarialAEC_SU}. A cycle-consistency loss was used in~\cite{gao2021cycu_neT_SU} to guide the reconstruction of two cascaded AECs in HU.
Self-supervised learning has also been integrated into the design of AEC strategies for HU~\cite{deshpande2021practicalDeepLearningUnmixing}.
Moreover, non-AEC approaches were also proposed for HU, such as parametrizations of the abundances using untrained deep prior models~\cite{rasti2021HU_deepImagePrior} or designing NNs for unmixing by unfolding optimization algorithms such as the ADMM~\cite{zhou2021ADMM_SU_networks} or sparse regression methods~\cite{qian2020modelInspiredNNs_SU}. Differently from approaches such as unfolded NNs, the proposed method is built upon a statistical model of the mixing and unmixing process, where some of the PDFs are parametrized by learnable models.

\section{Proposed method}
\label{sec:proposed}

\paragraph*{\textbf{Definitions}} 
Let us consider a dataset with $N_U$ unlabeled HI pixels $\calD_U=\{\by_1,\ldots,\by_{N_U}\}$. We also consider a supervised dataset $\calD_S=\{(\by_1,\ba_1,\bM_1),\ldots,(\by_{N_S},\ba_{N_S},\bM_{N_S})\}$ with $N_S$ labeled pixels (i.e., with its corresponding EMs and abundances), which will be generated directly from $\calD_U$ based on a self-supervised learning strategy that is described in detail in Section~\ref{sec:self_supervised_train}. We also denote by $\calD=\calD_U\bigcup\calD_S$ the full dataset. Functions without accent (e.g., $\bx$) belong to the mixing model, while functions with the \textit{breve} accent (e.g., $\breve{\bx}$) belong to the inference model. To simplify the notation, the pixel index will be omitted when possible.

\paragraph*{\textbf{Section overview}} 
In the following, we first describe the proposed mixing model in Section~\ref{sec:generative_model}. Then, the unmixing problem is formulated in a disentangled variational inference framework in Section~\ref{sec:unmixing_model}. The cost function for the semi-supervised learning objective is presented in~\ref{sec:cost_function}. Finally, the optimization of the proposed cost function is addressed in Section~\ref{sec:reparametrization}. 
An overview of the proposed framework is given in Figure~\ref{fig:overview}, which illustrates the interplay between the different parts of the statistical model representing the mixing (i.e., generative) and unmixing (i.e., inference) process, the generation of the training data, and the different terms in the loss function which is optimized to perform unmixing.

\subsection{The mixing model}
\label{sec:generative_model}

\paragraph*{\textbf{Abundance prior}} 
When assigning a statistical distribution to the abundances, it is very important to account for the physical constraints. 
The Dirichlet distribution is a natural choice for modeling abundance vectors since it enforces the non-negativity and sum-to-one constraints of the model, thus, being appropriate to represent fractions.
In this work we consider a flat Dirichlet distribution:
\begin{align}
    p(\ba) = \Dir(\ba;\cb{1}_P) \,,
    \label{eq:abundance_prior}
\end{align}
where $\cb{1}_P\in\amsmathbb{R}^{P}$ is a $P$-dimensional vector of ones that contains its concentration parameters of the distribution. The flatness indicates the lack of prior knowledge over the abundances other than its physical constraints. The Dirichlet distribution properties made it a popular choice of prior for Bayesian HU strategies~\cite{Halimi_IEEE_TIP_2015,eches2012adaptiveMRFunmixing,amiri2019bayesianSUdirichletSparse}, and has also been used in conjunction with a Markov transition model to represent the abundances multitemporal HU~\cite{bhatt2018multitemporalUnmixingDirichletMarkovAbundances}.

\paragraph*{\textbf{Endmember model}} 
Several models have been proposed to account for EM variability by representing the spectral signatures of the materials in each pixel using deterministic or statistical models. Deterministic models include the use of additive~\cite{Thouvenin_IEEE_TSP_2016_PLMM} or multiplicative~\cite{drumetz2016blindUnmixingELMM,imbiriba2018glmm,borsoi2020multitemporalUKalmanEM} perturbations of reference EM signatures, or a combination thereof~\cite{hong2019augmentedLMMvariability}.
The Gaussian~\cite{eismann2004otherNCMearly}, Beta~\cite{du2014spatialBetaCompositional} or Gaussian mixtures~\cite{zhou2018variabilityGaussianMixtureModel} distributions have been considered as statistical models. However, in practice EM variability can be very complex and specifying a distribution $p(\bM)$ directly is very difficult. Moreover, these models do not explore an important property of spectral variability. Since the spectrum of most materials is a function of a small number of physico-chemical parameters~\cite{HapkeBook1993,jacquemoud1990PROSPECTmodelLeaf}, the EM signatures are supported on a low-dimensional submanifold of the high-dimensional spectral space. %

In this work, we consider a deep generative EM model to provide a flexible representation of EM signatures while accounting for their low intrinsic dimension~\cite{borsoi2019deepGun}. Specifically, we model $\bM$ as a random variable that may be arbitrary but which follows a Gaussian distribution when conditioned on a set of low-dimensional latent variables $\bZ=[\bz_1,\ldots,\bz_P]$, $\bz_k\in\amsmathbb{R}^H$ (i.e., $p(\bM|\bZ)$ is Gaussian). The latent variables $\bZ$ control the variability of the spectral signatures, their dimension $H$ being related to the flexibility of the model to represent the EMs in an HI. Results in~\cite{borsoi2019deepGun} indicate that small values of $H$ are sufficient for obtaining good unmixing results and represent EM variability well. In this way, although the conditional distribution is tractable, the marginal PDF $p(\bM)=\int p(\bM|\bZ)p(\bZ)d\bZ$ can be arbitrary, rendering the model very flexible. Nonetheless, such a marginalization will not be necessary; instead, the latent variables $\bZ$ become an additional parameter that will be computed during inference. Considering the different EMs to be conditionally independent, the model becomes:
\begin{align}
    p_{\theta}(\bM|\bZ) 
    &= \prod_{k=1}^P p_{\theta}(\bm_k|\bz_k) \,,
\end{align}
with $\bm_k\in\amsmathbb{R}^L$ being the $k$-th column of $\bM$, and
\begin{align}
    p_{\theta}(\bm_k|\bz_k) &= \calN\big(\bm_k;\bmu_{\theta}^{m,k}(\bz_k),\diag(\bsigma_{\theta}^{m,k}(\bz_k)) \big) \,.
    \label{eq:p_m_z_gen}
\end{align}
Here $\calN(\bx;\bmu,\bSigma)$ denotes a Gaussian distribution of variable $\bx$ with mean $\bmu$ and covariance matrix $\bSigma$. Functions $\bmu_{\theta}^{m,k}(\bz_k)$ and $\diag(\bsigma_{\theta}^{m,k}(\bz_k))$ return the mean and (diagonal) covariance matrix of the distribution, and $\theta$ is the set of parameters of the generative model. We assign a prior for $\bZ$ as
\begin{align}
    p(\bZ) &= \prod_{k=1}^P p(\bz_k) \,, \,\,\,\,\,\,
    p(\bz_k) = \calN(\bz_k;\cb{0},\bI) \,.
\end{align}
Note that the number of EMs, $P$, is assumed to be known a priori. In practice, it can be estimated from the HI using virtual dimensionality estimation techniques~\cite{chang2018reviewVirtualDimensionalityHyperspectral}. However, special care has to be taken to avoid overestimating the number of EMs in the scene, as explained in~\cite{vijayashekhar2020virtualDimensionalityHyperspectralType1}.

\paragraph*{\textbf{Mixing model}}
Several models have been proposed to represent both macroscopic and intimate nonlinear mixtures, such as the BLMM and Hapke models~\cite{Dobigeon-2014-ID322,HapkeBook1993}. However, due
to the complexity of the mixing process, specifying a precise model can be difficult. This has motivated the development of HU method based on nonparametric models where the nonlinearity is learned from the data, such as in kernel-based methods~\cite{Chen-2013-ID321,borsoi2020BMUAN} and nonlinear AEC networks~\cite{wang2019AECnlin,hong2021egu_net,li2021modelBasedAECsSU}.
In this work, we consider a decomposition of the nonlinear mixing process as the sum of a linear term (the LMM) and a nonparametric nonlinear component~\cite{close2012usingPhysicsBasedMacroMicroUnmixing,Chen-2013-ID321,zhao2019AECnlin,li2021modelBasedAECsSU}:
\begin{align}
    \by = \bM\ba+\bmu_{\theta}^{y}(\ba,\bM) + \be \,,
    \label{eq:mixing_model}
\end{align}
where $\bmu_{\theta}^{y}$ denotes the nonlinear contribution.
This model aims at representing both macroscopic and intimate mixtures~\cite{close2012usingPhysicsBasedMacroMicroUnmixing}, with the linear term representing the main contribution of macroscopic interactions, and the nonlinear term representing the effects of multiple scattering and intimate mixtures. Its nonparametric form does not make direct assumptions about the type of mixture in an HI, at the cost of having to be learned. Moreover,
the amount of nonlinearity can be controlled by penalizing the norm of $\bmu_{\theta}^{y}$ during the learning process, which becomes close to the LMM when $\bmu_{\theta}^{y}$ is small.

Considering the noise $\be\in\amsmathbb{R}^L$ to be independent, white and Gaussian, the data likelihood becomes:
\begin{align}
    p_{\theta}(\by|\ba,\bM) = \calN\big(\by;\bM\ba+\bmu_{\theta}^{y}(\ba,\bM),\sigma_{\theta}^{y}\bI\big) \,,
    \label{eq:pixel_pdf}
\end{align}
where $\sigma_{\theta}^{y}\in\amsmathbb{R}_+$ is the noise variance in each band.

Since we assume that the abundances, the EMs and the noise are independent, this leads to the following factorization for the joint distribution:
\begin{align}
    \label{eq:generative_fact}
    p_{\theta}(\by,\ba,\bM,\bZ) = p_{\theta}(\by|\ba,\bM) p_{\theta}(\ba) p_{\theta}(\bM|\bZ) p(\bZ) \,.
\end{align}

\subsection{The unmixing problem}
\label{sec:unmixing_model}

The unmixing problem consists of finding the posterior distribution $p(\ba,\bM,\bZ|\by)$. However, due to the choice of distributions in~\eqref{eq:generative_fact} computing the posterior analytically is intractable, and some approximations have to be performed.
To this end, we use a variational approximation, which consists of specifying a parametric distribution $q_{\phi}(\ba,\bM,\bZ|\by)$ from a sufficiently rich family, and finding the set of parameters $\phi$ which makes it as close as possible to the true posterior distribution $p(\ba,\bM,\bZ|\by)$~\cite{bishop2006patternRecBook}. We consider the following factorization for this approximation:
\begin{align}
    q_{\phi}(\ba,\bM,\bZ|\by) 
    & = q_{\phi}(\ba|\by,\bM,\bZ)q_{\phi}(\bM|\bZ,\by) q_{\phi}(\bZ|\by) .
    \label{eq:sel_posterior_fact}
\end{align}

Moreover, we also consider disentanglement learning within a statistical framework, simplifying the model by separating latent variables through statistical independence assumptions~\cite{siddharth2017disentangledVAEs_NIPS}. Thus, we assume
that $\ba$ is independent of $\bZ$ conditioned on $\by$ and $\bM$, and that $\bM$ is independent of $\by$ conditioned on $\bZ$, that is:
\begin{align}
    q_{\phi}(\ba|\by,\bM,\bZ) & = q_{\phi}(\ba|\by,\bM) \,,
    \label{eq:disentangle_a}
    \\
    q_{\phi}(\bM|\bZ,\by) & = q_{\phi}(\bM|\bZ) \,.
    \label{eq:disentangle_m}
\end{align}
Through these two assumptions, $\ba$ is disentangled from $\bZ$, and $\bM$ is disentangled from $\by$. In the following, we define specific forms for each of those conditional PDFs.

For $q_{\phi}(\ba|\by,\bM)$, we consider another Dirichlet distribution:
\begin{align}
    q_{\phi}(\ba|\by,\bM) =  \Dir(\ba;\breve{\bgamma}_{\phi}^a(\by,\bM)) \,,
    \label{eq:var_post_abundances}
\end{align}
where function $\breve{\bgamma}_{\phi}^a$ computes its concentration parameters.

Most works (such as AECs) consider ``black-box'' models to compute the abundances from the mixed pixels. However, it is important to incorporate prior knowledge about the model to improve the interpretability of the results when parametrizing the abundances posterior distribution in~\eqref{eq:var_post_abundances}. Specifically, it has been shown that for autoencoder networks, the amount of nonlinearity in the mixing model and in the inference model are closely connected~\cite{li2021modelBasedAECsSU} in such a way that it is helpful to also decompose the encoder into a linear and nonlinear parts, so that the amount of nonlinearity in both parts of the model can be controlled.
Thus, we propose to write $\breve{\bgamma}_{\phi}^a(\by,\bM)$ as:
\begin{align}
    \breve{\bgamma}_{\phi}^a(\by,\bM) = s_{\text{ReLU}}\left(\breve{\bgamma}_{\phi}^{a,{\rm lin}}(\by,\bM) + \breve{\bgamma}_{\phi}^{a,{\rm nlin}}(\by,\bM) \right) \,,
    \label{eq:abundances_posterior_twoStream}
\end{align}
where $\breve{\bgamma}_{\phi}^{a,{\rm lin}}(\by,\bM)$ is a (piecewise) linear function which estimates the abundance mean and uncertainty based on a parametric model architecture which can also promote abundance sparsity effectively (as will be discussed in the following), while $\breve{\bgamma}_{\phi}^{a,{\rm nlin}}(\by,\bM)$ is a nonparametric function that is able to compensate other nonlinearities in the model. Function $s_{\text{ReLU}}(\bx)=\max(\cb{0},\bx)$ is the ReLU activation, which is used to ensure the nonnegativity of $\breve{\bgamma}_{\phi}^a(\by,\bM)$. Thus, we can use regularizations to explicitly control the contribution of the nonparametric term $\breve{\bgamma}_{\phi}^{a,{\rm nlin}}(\by,\bM)$ in the abundance estimates.

Distribution $q_{\phi}(\bZ|\by)$, is assumed to factorize as $q_{\phi}(\bZ|\by)=\prod_{k=1}^P q_{\phi}(\bz_k|\by)$, where each $q_{\phi}(\bz_k|\by)$ follows a Gaussian distribution:
\begin{align}
    q_{\phi}(\bz_k|\by) = \calN\big(\bz_k;\breve{\bmu}_{\phi}^{z,k}(\by),\diag(\breve{\bsigma}_{\phi}^{z,k}(\by))\big) \,,
    \label{eq:q_z_y}
\end{align}
in which functions $\breve{\bmu}_{\phi}^{z,k}$ and $\breve{\bsigma}_{\phi}^{z,k}$ compute its mean and the elements of its diagonal covariance matrix, respectively.

We also assume that that $q_{\phi}(\bM|\bZ)$ can be factorized as $q_{\phi}(\bM|\bZ)=\prod_{k=1}^P q_{\phi}(\bm_k|\bz_k)$, with:
\begin{align}
    q_{\phi}(\bm_k|\bz_k) = \calN\big(\bm_k; \breve{\bmu}_{\phi}^{m,k}(\bz_k),\diag(\breve{\bsigma}_{\phi}^{m,k}(\bz_k)) \big) \,.
    \label{eq:q_m_z}
\end{align}

\paragraph*{\textbf{Sharing parameters}} To simplify the inference process, we consider the same form for the EM conditional distribution in both~\eqref{eq:p_m_z_gen} and~\eqref{eq:q_m_z}. More precisely, we use $\breve{\bmu}_{\phi}^{m,k}(\bz_k)=\bmu_{\theta}^{m,k}(\bz_k)$ and $\breve{\bsigma}_{\phi}^{m,k}(\bz_k)=\bsigma_{\theta}^{m,k}(\bz_k)$, therefore making $q_{\phi}(\bM|\bZ)=p_{\theta}(\bM|\bZ)$. 
This is also known as ``bottom-up'' inference~\cite{sonderby2016trainLadderVAEs},
and allows for parameter sharing to reduce the complexity of the model.

\begin{figure}
    \centering
    \includegraphics[width=0.65\linewidth]{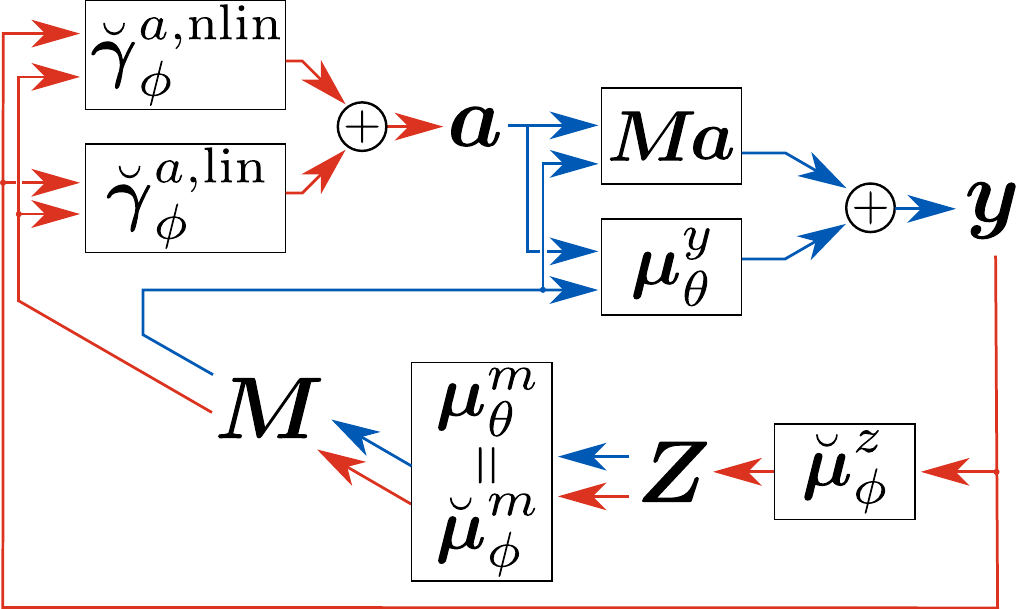}
    \vspace{-0.2cm}
    \caption{Overview of the dependencies between the variables in the proposed generative and inference (unmixing) models.
    Boxes denote functions, while the blue and red lines/arrows denote the data generating model and the inference model, respectively. \textit{Data generating model}: A latent variable $\bZ$ generates the EM matrix $\bM$, which together with the abundances $\ba$ generates the mixed pixel by an additive linear/nonlinear model. \textit{Inference model:} The learned model maps from the observed pixels to the latent EM representation $\bZ$, and from these to the estimated abundances using a ``bottom-up'' inference strategy and a two-stream neural network, respectively.}
    
    \label{fig:illustrative_diagram}
\end{figure}

An illustration of the relationship between the different random variables, $\ba,\bZ,\bM$ and $\by$, in both the generative and inference models according to the selected factorizations of the PDFs and their parametrizations can be seen in Figure~\ref{fig:illustrative_diagram}.

\subsection{Objective function}
\label{sec:cost_function}

To learn the parameters of the model and perform HU, we maximize a cost function composed of three terms, defined as:
\begin{align}
    \calL_{\rm T}(\theta,\phi;\calD) = \calL_{\rm Data}(\theta,\phi;\calD)  - \calL_{\rm Sparse}(\theta,\phi;\calD) - \calR(\theta,\phi) \,.
    \label{eq:cost_function}
\end{align}
The first term attempts to fit the model to the semi-supervised training dataset $\calD$, and aims to maximize the data likelihood $\log p(\calD)$. The second term aims to promote sparse abundance estimates, while the last term is a data-independent regularization on the model parameters.

\subsubsection{\textbf{Term $\calL_{\rm Data}(\theta,\phi;\calD)$}}

The main criterion used to learn the parameters of the model is to maximize the likelihood of the data:
\begin{align}
    \log p(\calD) = \sum_{\by\in\calD_U} \log p(\by) + \sum_{(\by,\ba,\bM)\in\calD_S} \log p(\by,\ba,\bM) \,,
    \label{eq:likelihood_1}
\end{align}
where the first term in the right hand side of~\eqref{eq:likelihood_1} consists in the data likelihood for the unlabeled dataset $\calD_U$, while the second term is the data likelihood for the labeled dataset $\calD_S$ (i.e., in which variables $\ba$ and $\bM$ are treated as fixed/observed, and thus their likelihood is also maximized).

However, maximizing the expression in~\eqref{eq:likelihood_1} is intractable because it involves marginalizing different variables in the joint distribution~\eqref{eq:generative_fact}, which is a common issue in latent variable models~\cite{barber2012ML_book}.
This problem can be circumvented by maximizing the so-called \textit{evidence lower bound} (ELBO) of each term in~\eqref{eq:likelihood_1}:
\begin{gather}
    \sum_{\by\in\calD_U} \log p(\by) \geq L_U(\theta,\phi;\calD_U) \,,
    \\
    \sum_{(\by,\ba,\bM)\in\calD_S} \log p(\by,\ba,\bM) \geq L_S(\theta,\phi;\calD_S) \,,
\end{gather}
where $L_U(\theta,\phi;\calD_U)$ and $L_S(\theta,\phi;\calD_S)$ are given by~\cite{siddharth2017disentangledVAEs_NIPS,kingma2014semiSupVAEs}:
\begin{align}
    L_{U}(\theta,\phi; & \calD_U) = \sum_{\by\in\calD_U} \Ex_{q_{\phi}(\ba,\bM,\bZ|\by)} \Big[\log p_{\theta}(\by|\ba,\bM,\bZ) 
    \nonumber \\
    & + \log p_{\theta}(\bZ,\ba,\bM) -\log q_{\phi}(\ba,\bM,\bZ|\by)  \Big] \,,
    \label{eq:dataTerm_unsup}
\end{align}
and
\begin{align}
    L_{S}(\theta,\phi;&\calD_S)  = \sum_{(\by,\ba,\bM)\in\calD_S} \Ex_{q_{\phi}(\bZ|\by,\ba,\bM)} \Big[\log p_{\theta}(\by|\ba,\bM,\bZ) 
    \nonumber \\
    & + \log p_{\theta}(\bZ,\ba,\bM) -\log q_{\phi}(\bZ|\by,\ba,\bM)  \Big] \,.
    \label{eq:dataTerm_sup}
\end{align}
Moreover, it can be shown that maximizing the ELBO equivalently minimizes the KL divergence between the variational posterior $q_{\phi}(\ba,\bM,\bZ|\by)$ and the (unknown) true posterior $p_{\theta}(\ba,\bM,\bZ|\by)$~\cite{kingma2019introductionVAEs}.

The cost function to be maximized is a combination of the lower bounds $L_U$ and $L_S$, and an additional regularization term~\cite{siddharth2017disentangledVAEs_NIPS,kingma2014semiSupVAEs}:
\begin{align}
    \calL_{\rm Data}(\theta,\phi;\calD) {}={} & L_U(\theta,\phi;\calD_U) + \lambda \Big[L_S(\theta,\phi;\calD_S)
    \nonumber \\
    & + \beta \sum_{(\by,\ba,\bM)\in\calD_S} \log q_{\phi}(\ba,\bM|\by) \Big] \,,
    \label{eq:dataTerm_initial}
\end{align}
where parameter $\lambda\in\amsmathbb{R}_+$ is used to balance the contribution of the supervised and unsupervised terms in the cost function. This is especially important when the size of the supervised dataset is much smaller than that of the unsupervised one~\cite{siddharth2017disentangledVAEs_NIPS}. The additional term $\log q_{\phi}(\ba,\bM|\by)$, weighted by $\beta\in\amsmathbb{R}_+$, is a regularization added to the cost function to promote models in which the labeled data (abundances and EMs) have high likelihood in the posterior, conditioned on the pixels~\cite{kingma2014semiSupVAEs}.

\subsubsection{\textbf{Term $\calL_{\rm Sparse}(\theta,\phi;\calD)$}}
An important property of many HIs is that while many materials may be present in the whole scene, only a small subset of them are usually present in each pixel. This leads the resulting abundance vectors to be sparse, with only a few of its elements being significantly far from zero~\cite{iordache2011sunsal}.
This property can be used to improve the conditioning of the unmixing problem in the form of regularization strategies. 
In this work, we penalize the concentration parameters of the abundance posterior distribution~\eqref{eq:var_post_abundances} to encourage sparse abundance reconstructions:
\begin{align}
    \calL_{\rm Sparse}(\theta,\phi;\calD) = \tau \sum_{(\by,\ba,\bM)\in\calD_S}
    \big\|\breve{\bgamma}_{\phi}^a(\by,\bM)\big\|_{1/2} \quad
    \nonumber \\
    + \tau \sum_{\by\in\calD_U} \Ex_{q_{\phi}(\bM|\bZ)q_{\phi}(\bZ|\by)}\Big\{\|\breve{\bgamma}_{\phi}^a(\by,\bM)\|_{1/2}\Big\} \,,
\end{align}
where the averages of the L$_{1/2}$ semi-norm of the Dirichlet concentration parameters are performed with respect to the supervised (first term) and unsupervised (second term) data portions, and $\tau\in\amsmathbb{R}_+$ is a parameter controlling the regularization. 
Furthermore, we considered the L$_{1/2}$ norm penalization as it is a more effective measure of sparseness when compared to the traditionally used L$_1$ norm penalty~\cite{qian2011unmixing_L12_NMF}.

\subsubsection{\textbf{Term $\calR(\theta,\phi)$}}

To control the nonlinear contributions and retain the interpretability of both the encoder~\eqref{eq:mixing_model} and decoder~\eqref{eq:abundances_posterior_twoStream} models, we add as a penalization term
\begin{align}
    \calR(\theta,\phi) = \varsigma_1\big\|\bmu_{\theta}^{y}\big\|_{\rm FNN} + \varsigma_2\big\|\breve{\bgamma}_{\phi}^{a,{\rm nlin}}\big\|_{\rm FNN} \,,
    \label{eq:regularization_params}
\end{align}
where $\|f\|_{\rm FNN}$ denotes the norm of some feedforward neural network $f$, defined as the sum of the norms of its parameters at all layers. For instance, if $f$ represents a multilayer preceptron (MLP) with weights $\bW_{\!\ell}$ and biases $\bb_{\ell}$ at layer $\ell$, the norm is computed as $\|f\|_{\rm FNN}=\sum_{\rm \ell} \|\bW_{\!\ell}\|_F+\|\bb_{\ell}\|_2$, with $\|\cdot\|_F$ denoting the Frobenius norm. The first regularization term in~\eqref{eq:regularization_params} penalizes the amount of nonlinearity in the mixing model~\eqref{eq:mixing_model}, while the second term penalizes the contribution of the nonparametric component in the abundance estimates in~\eqref{eq:abundances_posterior_twoStream}. Constants $\varsigma_1,\varsigma_2\in\amsmathbb{R}_+$ are regularization parameters.

\subsection{Optimizing the cost function}
\label{sec:reparametrization}

Due to the nature of the probabilistic model considered in this work, directly maximizing the cost function $\calL_{\rm T}(\theta,\phi;\calD)$ in~\eqref{eq:cost_function} is not trivial. This occurs because it involves some conditional distributions which are not available in the model factorization devised in Section~\ref{sec:unmixing_model}. Moreover, it also requires optimizing expectations with respect to both Gaussian and Dirichlet distributions depending on the model parameters. We address these problems in the following.

\subsubsection{\textbf{Rewriting the unavailable distributions}}
The first difficulty comes from the cost function $\mathcal{L}_{\text{Data}}(\theta,\phi;\calD)$ in~\eqref{eq:dataTerm_initial}. 
First, it requires sampling from and evaluating $q_{\phi}(\bZ|\by,\ba,\bM)$ (see~\eqref{eq:dataTerm_sup}), which is not available under the selected posterior factorization in~\eqref{eq:sel_posterior_fact}. Similarly, we also do not have access to the marginal posterior $q_{\phi}(\ba,\bM|\by)$.
To circumvent the first issue, we can proceed as in~\cite{siddharth2017disentangledVAEs_NIPS} and rewrite the last two terms of $\calL_{\rm Data}(\theta,\phi;\calD)$ as:
\begin{align}
    & L_{S}(\theta,\phi;\calD_S) + \beta \sum_{(\by,\ba,\bM)\in\calD_S} \log q_{\phi}(\ba,\bM|\by) 
    \nonumber \\
    & =
    \sum_{(\by,\ba,\bM)\in\calD_S} \Big[ \Ex_{q_{\phi}(\bZ|\by,\ba,\bM)} \Big\{\log p_{\theta}(\by|\ba,\bM,\bZ) 
    \nonumber \\
    & + \log p_{\theta}(\ba,\bM,\bZ) -\log q_{\phi}(\ba,\bM,\bZ|\by) \Big\}
    \nonumber \\
    & + (1+\beta) \log q_{\phi}(\ba,\bM|\by) \Big] \,,
    \label{eq:semisub_reworked_1}
\end{align}
where we added and subtracted the term $\log q_{\phi}(\ba,\bM|\by)$. Note that the variational posterior inside the expectation is now in the same form as~\eqref{eq:sel_posterior_fact}.

However, this still requires optimizing an expectation taken with respect to the unavailable $q_{\phi}(\bZ|\by,\ba,\bM)$. This can be approximated using importance sampling~\cite{siddharth2017disentangledVAEs_NIPS}, which allows us to approximate the expectation w.r.t. this term by sampling from $\bZ^{(i)}\sim q_{\phi}(\bZ|\by)$ (for each datapoint $\by\in\calD_U$), which is available under the factorization~\eqref{eq:sel_posterior_fact}. Using self-normalized importance sampling with $K$ samples, we can write the expectation in the right hand side of~\eqref{eq:semisub_reworked_1} as:
\begin{align}
    & \Ex_{q_{\phi}(\bZ|\by,\ba,\bM)} \big\{\log p_{\theta}(\by,\ba,\bM,\bZ) -\log q_{\phi}(\bZ,\ba,\bM|\by) \big\}
    \nonumber \\
    & = \Ex_{q_{\phi}(\bZ|\by)} \bigg\{ \frac{q_{\phi}(\bZ|\by,\ba,\bM)}{q_{\phi}(\bZ|\by)} \bigg[\log\frac{p_{\theta}(\by,\ba,\bM,\bZ)}{q_{\phi}(\bZ,\ba,\bM|\by)} \bigg] \bigg\}
    \nonumber \\
    & = \Ex_{q_{\phi}(\bZ|\by)} \bigg\{ \frac{q_{\phi}(\bZ|\by,\ba,\bM)q_{\phi}(\bM,\ba|\by)}{q_{\phi}(\bZ|\by)q_{\phi}(\bM,\ba|\by) } \nonumber\\ & \hspace{4cm}
    \times \bigg[\log \frac{p_{\theta}(\by,\ba,\bM,\bZ)}{q_{\phi}(\bZ,\ba,\bM|\by)} \bigg] \bigg\}
    \nonumber \\
    & = \Ex_{q_{\phi}(\bZ|\by)} \bigg\{ \frac{q_{\phi}(\bM|\bZ)}{q_{\phi}(\bM|\by)} \bigg[ \log \frac{p_{\theta}(\by,\ba,\bM,\bZ)}{q_{\phi}(\bZ,\ba,\bM|\by)} \bigg] \bigg\}
    \nonumber \\
    & \simeq \sum_{i=1}^K \frac{\omega^{(i)}}{\Omega} \log \frac{p_{\theta}(\by,\ba,\bM,\bZ^{(i)})}{q_{\phi}(\bZ^{(i)},\ba,\bM|\by)} \,,
    \label{eq:dataTerm_supterm_impSamp}
\end{align}
where~\eqref{eq:disentangle_a} and~\eqref{eq:disentangle_m} were used to obtain the third equality. Variable $\bZ^{(i)}$ is sampled as $\bZ^{(i)}\sim q_{\phi}(\bZ|\by)$, and the importance weights $\omega^{(i)}$ and their normalization factor $\Omega$ are obtained as
\begin{align}
    \frac{q_{\phi}(\bM|\bZ^{(i)})}{q_{\phi}(\bM|\by)}
    &\,\,\propto\,\, q_{\phi}(\bM|\bZ^{(i)}) 
    \triangleq \omega^{(i)} \,,
    \label{eq:importance_samp_w}
    \\
    \Omega &= \sum_{i=1}^K \omega^{(i)} \,.
    \label{eq:importance_samp_Omega}
\end{align}
Note that since $q_{\phi}(\bM|\by)$ is constant in~\eqref{eq:importance_samp_w}, i.e., it does not depend on $i$, and is also present in $\Omega$ in~\eqref{eq:dataTerm_supterm_impSamp} it cancels in~\eqref{eq:importance_samp_Omega}. Therefore, $q_{\phi}(\bM|\by)$ (which is not available under the chosen posterior factorization) is not required.

Finally, to approximate term $\log q_{\phi}(\ba,\bM|\by)$ in~\eqref{eq:semisub_reworked_1}, note that it can be lower bounded as~\cite{siddharth2017disentangledVAEs_NIPS}:
\begin{align}
    \log q_{\phi}(\ba, & \bM|\by) \geq \Ex_{q_{\phi}(\bZ|\by)} \bigg\{\log\frac{q_{\phi}(\ba,\bM,\bZ|\by)}{q_{\phi}(\bZ|\by)} \bigg\} 
    \label{eq:predictive_term_lowerbound} \\
    \nonumber
    & = \log q_{\phi}(\ba|\bM,\by) + \Ex_{q_{\phi}(\bZ|\by)} \Big\{\log q_{\phi}(\bM|\bZ) \Big\} \,.
\end{align}
Then, using a Monte Carlo estimator of the expectation in~\eqref{eq:predictive_term_lowerbound} with $K$ samples gives us:
\begin{align}
    \Ex_{q_{\phi}(\bZ|\by)} \Big\{\log q_{\phi}(\bM|\bZ) \Big\}
    & \simeq \frac{1}{K} \sum_{i=1}^K \log q_{\phi}(\bM|\bZ^{(i)})
    \nonumber\\
    & = \frac{1}{K} \sum_{i=1}^K \log \omega^{(i)} \,.
    \label{eq:dataTerm_reg_samp}
\end{align}
It can be seen that this approximation can be computed directly using the importance weights obtained in~\eqref{eq:importance_samp_w}.

To achieve the final cost function we re-write the terms in~\eqref{eq:cost_function} leveraging the previous results for the different terms. For the terms in $\calL_{\rm Data}(\theta,\phi;\calD)$ shown in~\eqref{eq:semisub_reworked_1} we used the lower bound in~\eqref{eq:predictive_term_lowerbound} and the importance sampling Monte Carlo approximation in~\eqref{eq:dataTerm_supterm_impSamp}. For the remaining $L_U(\theta,\phi;\calD_U)$ term and $\calL_{\rm Sparse}(\theta,\phi;\calD)$ we applied Monte Carlo approximations with $K_E$ samples to compute its expectations. The resulting cost function is shown in equation~\eqref{eq:data_term_cf_sampled}.

\begin{figure*}[ht]
    \centering
    {%
    \begin{align}
        \widehat{\calL}_{\rm T}(\theta,\phi;\calD) & \simeq \sum_{\by\in\calD_U} \frac{1}{K_E}\sum^{K_E}_{\substack{i=1 \\ \ba^{(i)},\bM^{(i)},\bZ^{(i)}\sim q_{\phi}(\ba|\bM,\by)q_{\phi}(\bM|\bZ)q_{\phi}(\bZ|\by) }} \bigg[\log \frac{p_{\theta}(\by|\ba^{(i)},\bM^{(i)}) p_{\theta}(\ba^{(i)}) p_{\theta}(\bM^{(i)}|\bZ^{(i)}) p(\bZ^{(i)})}{q_{\phi}(\ba^{(i)}|\bM^{(i)},\by)q_{\phi}(\bM^{(i)}|\bZ^{(i)}) q_{\phi}(\bZ^{(i)}|\by)} \bigg]
        \nonumber \\
        & + \lambda \sum_{(\by,\ba,\bM)\in\calD_S} \bigg[ \frac{1}{K} \sum^{K}_{\substack{j=1 \\ \bZ^{(j)}\sim q_{\phi}(\bZ|\by) }} \frac{q_{\phi}(\bM|\bZ^{(j)})}{\sum_{\ell=1}^{K} q_{\phi}(\bM|\bZ^{(\ell)})} \bigg[ \log \frac{p_{\theta}(\by|\ba,\bM) p_{\theta}(\ba) p_{\theta}(\bM|\bZ^{(j)}) p(\bZ^{(j)})}{q_{\phi}(\ba|\bM,\by)q_{\phi}(\bM|\bZ^{(j)}) q_{\phi}(\bZ^{(j)}|\by)} \bigg]
        \nonumber \\
        & + (1+\beta) \Big[\log q_{\phi}(\ba|\bM,\by) + \frac{1}{K} \sum^{K}_{\substack{m=1 \\ \bZ^{(m)}\sim q_{\phi}(\bZ|\by) }} \log q_{\phi}(\bM|\bZ^{(m)})\Big]
        \bigg]
        - \tau \sum_{(\by,\ba,\bM)\in\calD_S} 
        \big\|\breve{\bgamma}_{\phi}^a(\by,\bM)\big\|_{1/2}
        \nonumber \\
        & - \tau \sum_{\by\in\calD_U} \bigg[ \frac{1}{K_E}\sum^{K_E}_{\substack{i=1 \\ \bM^{(i)},\bZ^{(i)}\sim q_{\phi}(\bM|\bZ)q_{\phi}(\bZ|\by) }} \big\|\breve{\bgamma}_{\phi}^a(\by,\bM^{(i)})\big\|_{1/2}\bigg]
        - \calR(\theta,\phi) \,.
        \label{eq:data_term_cf_sampled} 
    \end{align}
    }
\vspace{-3ex}
\end{figure*}

\subsection{Reparametrization}

In order to optimize the cost function in~\eqref{eq:data_term_cf_sampled} with respect to the parameters $\theta$ and $\phi$, it is necessary to estimate the gradient of expressions involving variables $\bZ^{(i)}$, $\bM^{(i)}$ and $\ba^{(i)}$, which are sampled from the variational posterior distribution $q_{\phi}(\ba,\bM,\bZ|\by)$. However, since this distribution depends on the parameters $\phi$, it is necessary to account for the dependency of the samples $\bZ^{(i)}$, $\bM^{(i)}$ and $\ba^{(i)}$ on $\phi$ when computing the gradients.
To this end, we consider the reparametrization trick, which provides low-variance gradient estimates for such problems~\cite{schulman2015gradientEstimationPathDer}. In general, this is performed by writing the considered random variables (say, $\bx$) in terms of a distribution that does not depend on $\phi$, e.g., for a given function $f$ to be differentiated, we have:
\begin{align}
    f(\bx) = f(g(\bepsilon,\phi)) \,,
\end{align}
where $g$ is a function such that $x$ and $g(\bepsilon,\phi)$ have the same distribution, and $\bepsilon\sim p(\bepsilon)$ does not depend on $\phi$. Thus, the gradient $\partial f(\bx)/ \partial \phi$ can be computed straightforwardly and estimated using Monte Carlo sampling.

This strategy can be applied to reparametrize the samples from  $q_{\phi}(\bM|\bZ)$ and $q_{\phi}(\bZ|\by)$. Since the variables $\bZ$ and $\bM$ are conditionally Gaussian, the reparametrization trick is straightforward to apply (as in~\cite{kingma14VAEs}), resulting in:
\begin{align}
    \bZ^{(i)} & =  \big[\breve{\bmu}_{\phi}^{z,1}(\by),\ldots,\breve{\bmu}_{\phi}^{z,P}(\by)\big]
    \nonumber \\
    & \quad + \big[\breve{\bsigma}_{\phi}^{z,1}(\by),\ldots,\breve{\bsigma}_{\phi}^{z,P}(\by)\big] \odot\bXi_z^{(i)} \,,
    \\
    \bM^{(i)} &= \big[\breve{\bmu}_{\phi}^{m,1}(\bZ_1^{(i)}),\ldots,\breve{\bmu}_{\phi}^{m,P}(\bZ_P^{(i)})\big]
    \nonumber\\
    & \quad + \big[\breve{\bsigma}_{\phi}^{m,1}(\bZ_1^{(i)}),\ldots,\breve{\bsigma}_{\phi}^{m,P}(\bZ_P^{(i)})\big] \odot \bXi_m^{(i)} \,,
\end{align}
where $\odot$ is the Hadamard (i.e., elementwise) product and $\bXi_z^{(i)}$, $\bXi_m^{(i)}$ are matrices whose elements are i.i.d. and sampled from a Gaussian random variable with zero mean unity variance.

Since the last term $q_{\phi}(\ba|\by,\bM)$ is a Dirichlet distribution, the common reparametrization trick is not applicable. Various strategies have been considered to address this problem in deep variational inference, including approximating it as a Gaussian in the softmax basis~\cite{srivastava2017autoencodingDirSoftmax} or reparametrizing rejection samplers~\cite{naesseth2017rejectionSamplVAE}. 
Here we consider the so-called pathwise derivative estimator proposed in~\cite{jankowiak2018pathwiseDerivatives}, which allows us to estimate the derivative of a stochastic variable $\ba$ sampled from a Dirichlet distribution $\Dir(\ba;\bgamma)$ as:
\begin{align}
    \frac{\partial a_i}{\partial \gamma_j} = - \frac{\frac{\partial F_{\rm Beta}}{\partial \gamma_j}\big(a_j; \gamma_j, \sum_{k=1}^P\gamma_k-\gamma_j\big)}{f_{\rm Beta}\big(a_j; \gamma_j, \sum_{k=1}^P\gamma_k-\gamma_j\big)} \Big(\frac{\delta_{i-j}-a_i}{1-a_j}\Big) \,,
\end{align}
where $f_{\rm Beta}$ and $F_{\rm Beta}$ are the Beta probability and cumulative distribution functions, respectively, and $\delta_{i-j}$ the Kronecker delta function, satisfying $\delta_{i-j}=1$ if $i=j$ and zero otherwise (see~\cite{jankowiak2018pathwiseDerivatives} for more details). This expression can be used to compute the gradient of the cost function using the chain rule.

\subsection{On the interpretability of the proposed approach}

Much interest has been recently dedicated to interpretability in the field of machine learning~\cite{miller2019explanationAI,arrieta2020explainable}, and while a mathematical formalization of interpretability is difficult, a definition that is appropriate in this context is when it allows a user/human observer to understand the model and algorithm’s behavior~\cite{miller2019explanationAI,arrieta2020explainable}.
In the proposed method, interpretability mainly comes from the statistical modeling and residual network design.

First, the unmixing problem tackled in this work involves multiple nonidealities, such as nonlinearity and variability of the endmembers. By modeling both the generative and inference (unmixing) models as PDFs (in~\eqref{eq:generative_fact} and~\eqref{eq:sel_posterior_fact}) and employing adequate conditional independence assumptions on the different model variables (e.g., in~\eqref{eq:disentangle_a}
\eqref{eq:disentangle_m}), we are able to separately account for each of these effects, instead of having a single black box model. Moreover, the independence assumptions used in the inference model allow us to disentangle the influence of each of these effects in the recovered abundances~\cite{siddharth2017disentangledVAEs_NIPS}. This framework also leads to a principled training procedure based on semi-supervised learning, where the contribution of the data with and without labels (which might be extracted from the HI through self-supervised learning, as detailed later) is clear.
Another point is that the residual formulation of the mixing model, i.e., a linear mixing model augmented with a flexible nonlinear component, see~\eqref{eq:mixing_model}, copes with different types of mixtures, such as, e.g., the multiple mixture pixel model~\cite{close2012usingPhysicsBasedMacroMicroUnmixing} or BLMMs~\cite{Dobigeon-2014-ID322}.

\section{Neural network design and training}
\label{sec:nnets_details}

In this section, we detail different aspects of the method, consisting in: 1) the choices of model and NN architectures, 2) the construction of the supervised training data using a self-supervision strategy, and 3) the cost function optimization strategy.

\subsection{Model details and neural network architectures}

\subsubsection{\textbf{Endmember networks}}

For the EM networks, which is used in both the distribution $p_{\theta}(\bm_k|\bz_k)$ in the mixing model~\eqref{eq:p_m_z_gen} and $q_{\phi}(\bm_k|\bz_k)$ in the inference network~\eqref{eq:q_m_z}, we considered the following architectures. For the mean vectors $\breve{\bmu}_{\phi}^{m,k}(\bz_k)=\bmu_{\theta}^{m,k}(\bz_k)$, we considered a fully connected MLP as in~\cite{borsoi2019deepGun} with six layers containing $H$, $\max\big\{\lceil L/10 \rceil,\, H+1\big\}$, $\max\big\{\lceil L/4 \rceil,\, H+2\big\} + 3$, $\lceil 1.2\times L \rceil + 5$ and $L$ neurons. The ReLU activation function was used on the hidden layers and a sigmoid on the last layer.
For the covariances, for simplicity we used an EM-wise learnable constant $\sigma^{m,k}\in\amsmathbb{R}_+$ resulting in an isotropic model, which led to $\breve{\bsigma}_{\phi}^{m,k}(\bz_k)=\bsigma_{\theta}^{m,k}(\bz_k)=\sigma^{m,k}\cb{1}_L$, where $\cb{1}_L$ is an $L$-dimensional vector of ones.

\subsubsection{\textbf{Variability inference network}}

For the variational distributions $q_{\phi}(\bz_k|\by)$ in~\eqref{eq:q_z_y}, we considered two fully connected MLPs with partially shared parameters to compute the mean and diagonal covariance elements. For the mean $\breve{\bmu}_{\phi}^{z,k}(\by)$, we considered four fully connected layers with $L$, $5H$, $2H$ and $H$ neurons. For the diagonal covariance elements $\breve{\bsigma}_{\phi}^{z,k}(\by)$, we considered six layers with $L$, $5H$, $2H$, $2H$, $2H$ and $H$ neurons. The first two hidden layers were shared between both networks, and the ReLU activation function was used in all hidden layers, with a linear activation in the last layer.

\subsubsection{\textbf{Nonlinear mixing (decoder) network}} The measurement model $p_{\theta}(\by|\ba,\bM)$ in~\eqref{eq:pixel_pdf} was parametrized as follows. 
For the function $\bmu_{\theta}^{y}(\ba,\bM)$ representing the nonlinear contribution in the mixing model~\eqref{eq:mixing_model}, we consider fully connected MLP as used in~\cite{li2021modelBasedAECsSU}. It consisted of five fully connected layers, with $P(L+1)$, $PL$, $L$, $L$ and $L$ neurons. The ReLU activation was used in all hidden layers, and a linear activation was used in the last layer. %
The observation noise variance 
$\sigma_{\theta}^{y}$ in~\eqref{eq:pixel_pdf} is set as a positive learnable constant.

\subsubsection{\textbf{Abundance posterior (encoder) network}}
\label{eq:abundance_rec_nn}

As discussed previously, we consider a two-stream architecture in~\eqref{eq:abundances_posterior_twoStream} to estimate the abundance posterior distribution~\eqref{eq:var_post_abundances}. For the piecewise linear reconstruction function $\breve{\bgamma}_{\phi}^{a,{\rm lin}}(\by,\bM)$, we propose to consider an unrolled NN architecture that is suitable to sparse regression. 
Unrolling algorithms to create NNs consists of converting the iterative equations of traditional optimization algorithms (e.g., gradient descent or iterative shrinkage/thresholding) into NN layers. The different parameters of the algorithms, which are traditionally fixed (or come from a forward measurement model) become the trainable parameters of the NN~\cite{monga2021unrollingNNetsReview}. 
Such architectures have become very prominent in machine learning solution to various inverse problems due to their high interpretability~\cite{monga2021unrollingNNetsReview}.

In this work, we consider an architecture inspired by the LISTA method~\cite{ablin2019lista} with $M$ network layers, which unfolds the iterative shrinkage/thresholding algorithm for sparse regression. This leads to the following architecture, with the $m$-th hidden network layer, $m\in\{1,\ldots,M-1\}$, given by:
\begin{align}
    \bh^{(m+\frac{1}{2})} &= \bh^{(m)} - \eta^{(m)}  \bM^\top \big(\bM\bh^{(m)} - \by\big) \,,
    \\
    \bh^{(m+1)} &= s_{\text{ReLU}}\big(\bh^{(m+\frac{1}{2})} - \eta^{\text{sp}}\eta^{(m)} \cb{1}_P\big) \,,
\end{align}
where $\bh^{(m)}$ denotes the signal in the $m$-th later, $\eta^{(m)}\in\amsmathbb{R}_+$ is a learnable step size and $\eta^{\text{sp}}\in\amsmathbb{R}_+$ a parameter that controls sparsity of the reconstruction. The hidden layers can be interpreted as a gradient step of a least squares problem, followed by a combined shrinkage and projection to the nonnegative orthant. 
The output of the first layer is computed as $\bh^{(1)}=\bM^\dagger\by$, with $\bM^\dagger$ denoting the pseudoinverse of $\bM$. The last layer, computed as $\bh^{(M)}=\eta^{\text{unc}}\bh^{(M-1)}$, consists of a simple scaling of the result by a scalar parameter $\eta^{\text{unc}}\in\amsmathbb{R}_+$, which controls the spread of the Dirichlet distribution.
This approach is interesting since the learnable parameters are just the step sizes $\eta^{\text{unc}}$ and sparsity and uncertainty parameters $\eta^{\text{sp}}$, $\eta^{(m)}$, which maintains a high degree of interpretability in the architecture.

For the nonlinear part of the abundance posterior $\breve{\bgamma}_{\phi}^{a,{\rm nlin}}(\by,\bM)$, we considered an MLP with six fully connected layers containing $L$, $2L$, $\lfloor L/2\rceil$, $\lfloor L/4\rceil$, $4P$ and $P$ neurons, where $\lfloor\cdot\rceil$ denotes rounding to the nearest integer. The ReLU activation function was used in the hidden layers, and a linear activation in the last layer.

\subsection{Self-supervised training strategy}
\label{sec:self_supervised_train}

An important aspect of the semi-supervised framework is the need to obtain some amount of supervised training data. As done in previous works~\cite{borsoi2019deepGun,hong2021egu_net}, we propose to use a self-supervised strategy based on methods which extract different samples from EMs directly from an observed HI (i.e., in the form of pure pixels)~\cite{borsoi2020variabilityReview,somers2012automatedBundlesRansomSampling}. These methods have been originally proposed as image-based spectral library construction to address spectral variability in library-based HU algorithms without the requirement of laboratory or \textit{in situ} data collection. Nonetheless, these are also suitable strategies to generate training data for learning-based HU methods.

In this paper, we consider the strategy proposed in~\cite{borsoi2019deepGun}, to generate a supervised synthetic dataset $\calD_\text{S}$ directly from the HI being analyzed by following a two step procedure. In the first step, we create a dictionary $\calD_\text{ppx}$ of pure pixels. Pure pixels are obtained by extracting a predetermined number $N_{\text{ppx}}$ of pixels from the HI which are spectrally close (e.g., having the smallest spectral angles) to a set of reference EMs obtained using a traditional EM extraction algorithm (such as the VCA~\cite{Nascimento2005}).
The second step consists in an iterative algorithm for generating synthetic data that incorporates the variability existing within $\calD_\text{ppx}$. For this, at each iteration $k$ we generate tuples $\{(\by_{j},\ba_{j},\bM_k)\}_{j=1}^{P}$ where $\bM_k\in\amsmathbb{R}^{L\times P}$ is an EM matrix sampled from the pure pixel dictionary. For each $k$ we generate $P$ abundance vectors $\ba_j$, $j\in\{1,\ldots,P\}$, with the elements $a_{j,i}$ of the $j$-th vector satisfying $a_{j,i}=1$ if $i=p$ and $a_{j,i}=0$ otherwise (i.e., a one-hot encoding of each EM). Then, the mixed pixels are generated according to $\by_j=\bM_k\ba_j+\be$, with $\be$ being white Gaussian noise.

Note that since the synthetic abundances are one-hot encodings, this leads to a simple but effective means of generating training data, since nonlinear interactions (i.e., a nonlinear mixing model) don't need to be specified a priori due to the pixels being pure pixels.
Nevertheless, more general methods using a finer sampling of the unit simplex and different choices of mixing models can also be used to generate the training data at the expense of a higher amount of user supervision and possibly at a higher computation cost.

\subsection{Optimization}

To optimize the cost function, we considered the Adam stochastic optimization method~\cite{kingma2014adam}. We used a batch size of $16$ and trained the model for up to $30$ epochs, stopping when the relative increase in the value of the cost function $\widehat{\calL}_{\rm T}(\theta,\phi;\calD)$ between two epochs was smaller than $0.01$. %
The learning rate was initially set as $0.001$ and multiplied by $0.9$ after each epoch until the $10$th epoch using a learning rate scheduler, after which is was kept fixed.

\section{Experiments}
\label{sec:experiments}

\subsection{Experimental setup}

The proposed method is compared with the fully constrained least squares (FCLS) algorithm, with EMs extracted by the VCA~\cite{Nascimento2005}, with the PLMM~\cite{Thouvenin_IEEE_TSP_2016_PLMM}, ELMM~\cite{drumetz2016blindUnmixingELMM}, and GLMM~\cite{imbiriba2018glmm} algorithms, which account for EM variability, with the NDU~\cite{ammanouil2016nonlinear}, which considers a nonparametric nonlinear mixing model, and with recent deep learning-based strategies that address EM variability and nonlinearity, being DeepGUn~\cite{borsoi2019deepGun}, RBF-AEC~\cite{shahid2021unsupervisedSUautoencoder} and EGU-Net~\cite{hong2021egu_net}.
In each experiment the parameters of each algorithm were adjusted in order to obtain the best results.
All experiments were run in an AMD Ryzen~5 3500U portable computer with 6GB of RAM.

The proposed method was implemented in Pytorch. We consider $K_E=1$ sample for the Monte Carlo estimation of the expectations, and $K=5$ for the importance sampling. The supervised dataset $\calD_S$ was constructed using the self-supervised strategy detailed in Section~\ref{sec:self_supervised_train}, with VCA being used to extract the reference EM signatures, $N_{\text{ppx}}=100$ pure pixels from each EM being extracted to generate the dataset, and a signal-to-noise ratio (SNR) of 30dB was considered for the additive noise. The dimension of the EM latent space was set as $H=2$, and $M=11$ layers (i.e., ten hidden layers) were considered in the piecewise-linear abundance inference network detailed in Section~\ref{eq:abundance_rec_nn}. The remaining parameters of the method were manually selected within the following intervals: $\lambda\in[0.01,1000]$ with one value per decade; $\tau$ was either zero or selected in $\tau\in[0.001,1]$ with two values per decade; and the nonlinearity regularizations in the interval $\varsigma_1,\varsigma_2\in[10^{-4},10^5]$ with one value per decade.
For all methods, the number of EMs $P$ was assumed to be known a priori for the simulations with synthetic data. For the simulations with real data, we used the same values as in previous works that studied the same HIs. When such knowledge is not available, this number can be estimated from the HI~\cite{chang2018reviewVirtualDimensionalityHyperspectral}, although care must be taken to avoid overestimating the number of EMs in the scene~\cite{vijayashekhar2020virtualDimensionalityHyperspectralType1}.

To evaluate the results of the algorithms on the synthetic datasets, we considered the normalized mean squared error (NRMSE), defined as $\text{NRMSE}_{\bX} = \|\bX-\widehat{\bX}\|_F \big/ \|\bX\|_F$,
computed between a matrix $\bX$ and its estimated version $\widehat{\bX}$, where $\|\cdot\|_F$ denotes the Frobenius norm of a matrix or higher order tensor. We evaluated the NRMSE between the abundances ($\text{NRMSE}_{\bA}$) and between the EMs ($\text{NRMSE}_{\bM}$). We also evaluated the normalized image reconstruction error ($\text{NRMSE}_{\bY}$); however, we emphasize that a low values of $\text{NRMSE}_{\bY}$ do not imply good unmixing performance.

We also compute the spectral angle mapper (SAM) between the true and the estimated EMs at each pixel:
\begin{align}
    \text{SAM}_{\bM} = \frac{1}{N_U} \sum_{n=1}^{N_U} \sum_{j=1}^P \arccos \bigg( \frac{\bm_{n,j}^\top\widehat{\bm}_{n,j}}{\|\bm_{n,j}\|_2\|\widehat{\bm}_{n,j}\|_2} \bigg) \,.
\end{align}
where $\bm_{n,j}$ and $\widehat{\bm}_{n,j}$ are the true and estimated versions of the $j$-th EM in the $n$-th pixel, respectively.

\subsection{Synthetic data}

Two synthetic datasets with ground truth were considered, one containing nonlinearity in the mixing process and another containing EM variability. 
These two nonidealities are intruduced in separate datasets to have a more controlled experimental scenario, particularly since some of the compared methods were devised taking only nonlinearity or EM variability into account. However, in real datasets, especially over large scenes, these effects can occur concurrently.

\begin{figure*}
    \centering
    \includegraphics[height=0.375\pdfpageheight]{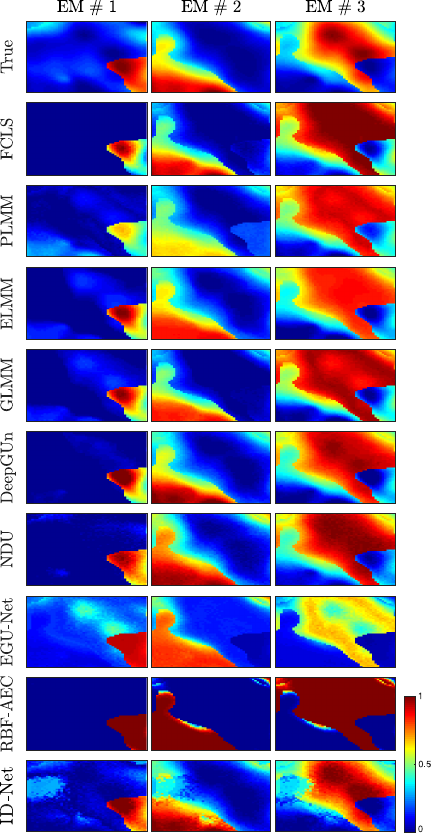}
    \hspace{0.1\linewidth}
    \includegraphics[height=0.375\pdfpageheight]{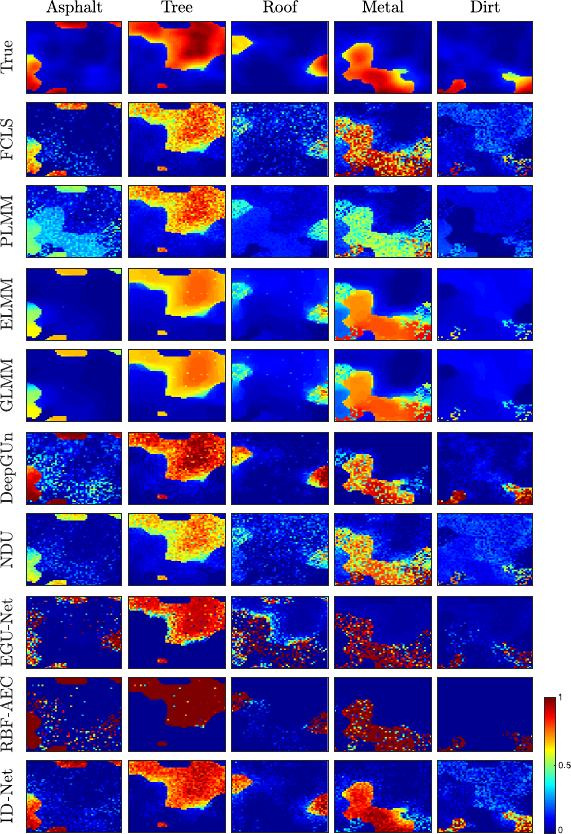}
    \vspace{-0.2cm}
    \caption{True and reconstructed abundance maps for the experiments with synthetic data for data cubes DC1 (left) and DC2 (right).}
    \label{fig:ex_synth_abunds}
\end{figure*}

\paragraph*{\textbf{Data with nonlinearity}}
To generate the first datacube (DC1), with $N_U=2500$ pixels, we considered synthetic abundance maps displayed in the first row of Figure~\ref{fig:ex_synth_abunds} (left) and $P=3$ EM signatures with $L=224$ spectral bands selected from the USGS Spectral Library.
The reflectance of each pixel was generated according to the BLMM $\by_n=\bM\ba_n+\sum_{i=1}^P\sum_{j=i+1}^Pa_{n,i}a_{n,j}\bm_i\odot\bm_j+\be_n$, with $\be_n$ being white Gaussian noise with an SNR of 30dB.

\paragraph*{\textbf{Data with endmember variability}}
To generate the second datacube (DC2), with $N_U=2500$ pixels, we considered synthetic abundance maps displayed in the first row of Figure~\ref{fig:ex_synth_abunds} (right). To incorporate EM variability in this dataset, sets of EM signatures from $P=5$ pure materials (roof, metal, dirt, tree and asphalt) with realistic variability were first manually extracted from a real HI. Then, to generate each pixel $\by_n$, spectral signatures for each EM were then randomly sampled from these sets and used in a generalized version of the LMM $\by_n=\bM_n\ba_n+\be_n$ with pixelwise EM matrices, where $\be_n$ was white Gaussian noise with an SNR of 30dB.

Note that the two synthetic datacubes incorporate nonidealities seen in practical scenes. The BLMM, used in DC1, has been experimentally shown to accurately describe multiple scattering in mixtures containing vegetation~\cite{somers2009nonlinearMixtureOrchards}. The different EM samples used to construct DC2 were manually extracted from a real HI, being representative of the variability encountered in a real image.

\paragraph*{\textbf{Discussion}}
The quantitative and visual results for both data cubes are presented in Table~\ref{tab:results_synthData} and Figure~\ref{fig:ex_synth_abunds}, respectively.
It can be seen that ID-Net obtained the best abundance estimation accuracy $\text{NRMSE}_{\bA}$ for both data cubes, with improvements of 39\% and 25\% over the second-best results. It also obtained a good EM estimation performance, with the best $\text{NRMSE}_{\bM}$ for DC1 and the best $\text{SAM}_{\bM}$ for DC2. 
The visual results of Figure~\ref{fig:ex_synth_abunds} show that although most methods were able to obtain a rough separation between the different materials in the scene, the reconstructions by ID-Net were the closest to the ground truth. This can be observed most clearly, for instance, in the abundance maps of the Metal EM in DC2.
The smallest reconstruction errors
$\text{NRMSE}_{\bY}$ were obtained by NDU, which employs a nonparametric model with many degrees of freedom, which can represent a pixel arbitrarily well. However, this does not imply better unmixing performance (i.e., better abundance and EM reconstructions).
The execution times of ID-Net, also shown in Table~\ref{tab:results_synthData}, are on the same order of magnitude of the more complex competing algorithms (such as PLMM, DeepGUn and NDU) but considerably higher than the faster ones (such as EGU-Net).

\begin{table}%
\footnotesize
\caption{Simulation results using synthetic data.}
\vspace{-0.2cm}
\centering
\renewcommand{\arraystretch}{1.3}
\setlength{\tabcolsep}{3.3pt}
\begin{tabular}{l|ccccr}
\bottomrule
& $\text{NRMSE}_{\bA}$ & $\text{NRMSE}_{\bM}$ & $\text{SAM}_{\bM}$ & $\text{NRMSE}_{\bY}$ & Time \\
\hline
&\multicolumn{5}{|c}{Data Cube 1 -- DC1} \\
\hline
FCLS	&	0.371	&	--	&	--	&	0.093	&	1.16	\\
PLMM	&	0.337	&	0.166	&	0.089	&	0.035	&	842.9	\\
ELMM	&	0.236	&	0.101	&	0.012	&	0.056	&	95.0	\\
GLMM	&	0.345	&	0.085	&	0.032	&	0.051	&	91.0	\\
DeepGUn	&	0.246	&	0.068	&	0.020	&	0.058	&	607.4	\\
NDU	&	0.277	&	--	&	--	&	0.034	&	1128.59	\\
EGU-Net	&	0.313	&	0.163	&	0.033	&	0.134	&	5.4	\\
RBF-AEC	&	0.458	&	0.129	&	0.067	&	0.068	&	82.0	\\
ID-Net	&	0.170	&	0.040	&	0.016	&	0.042	&	436.2	\\

\hline
&\multicolumn{5}{|c}{Data Cube 2 -- DC2}	\\
\hline
FCLS	&	0.511	&	--	&	--	&	0.275	&	1.1	\\
PLMM	&	0.507	&	0.625	&	0.487	&	0.048	&	1050.6	\\
ELMM	&	0.439	&	0.471	&	0.145	&	0.202	&	52.4	\\
GLMM	&	0.437	&	0.486	&	0.197	&	0.198	&	64.5	\\
DeepGUn	&	0.240	&	0.307	&	0.091	&	0.119	&	678.5	\\
NDU	&	0.516	&	--	&	--	&	0.0001	&	418.5	\\
EGU-Net	&	0.758	&	0.372	&	0.209	&	0.302	&	6.3	\\
RBF-AEC	&	0.514	&	0.292	&	0.134	&	0.145	&	78.3	\\
ID-Net	&	0.191	&	0.313	&	0.084	&	0.159	&	725.5	\\
\toprule		
\end{tabular}
\label{tab:results_synthData}
\end{table}

\begin{table}[htb]
\footnotesize
\caption{Simulation results using real data.}
\vspace{-0.2cm}
\centering
\renewcommand{\arraystretch}{1.35}
\setlength{\tabcolsep}{3pt}
\begin{tabular}{l|cr|cr|cr}
\bottomrule
& \multicolumn{2}{c|}{Samson} & \multicolumn{2}{c|}{Jasper Ridge} & \multicolumn{2}{c}{Cuprite}\\
\hline
& $\text{NRMSE}_{\bY}$ & Time  & $\text{NRMSE}_{\bY}$ & Time & $\text{NRMSE}_{\bY}$ & Time\\
\hline
FCLS	&		0.069	&	2.5	&		0.397	&	2.7	& 0.018 & 33.4 \\
PLMM	&		0.024	&	225.9	&		0.055	&	526.4	& 0.025 & 13885.9 \\
ELMM	&		0.012	&	64.9	&		0.028	&	138.1	& 0.017 & 3181.6 \\
GLMM	&		0.001	&	96.8	&		0.003	&	190.0	& 0.024 & 4743.7 \\
DeepGUn	&		0.075	&	676.4	&		0.111	&	1454.1	& 0.032 & 46685.8 \\
NDU	&		0.032	&	749.2	&		0.084	&	2138.7	& 0.016 & 8096.1 \\
EGU-Net	&		0.320	&	7.4	&		0.178	&	0.3	& 0.070 & 1479.3 \\
RBF-AEC	&		0.206	&	230.9	&		0.151	&	336.0	& 0.047 & 1298.0\\
ID-Net	&		0.057	&	1606.3	&		0.155	&	2209.4	& 0.143 & 4893.0\\
\toprule		
\end{tabular}
\label{tab:results_realData}
\end{table}

\begin{figure}
    \centering
    \begin{minipage}{0.32\linewidth}
    \includegraphics[height=0.09\pdfpageheight]{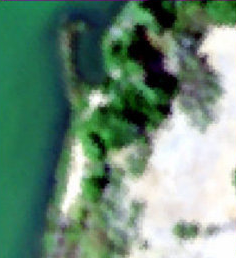}\\
    \centering \small Samson HI
    \end{minipage}
    \begin{minipage}{0.32\linewidth}
    \includegraphics[height=0.09\pdfpageheight]{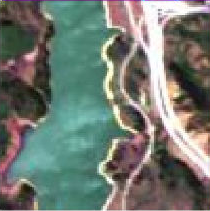}\\
    \centering \small Jasper Ridge HI
    \end{minipage}
    \begin{minipage}{0.32\linewidth}
    \includegraphics[height=0.09\pdfpageheight]{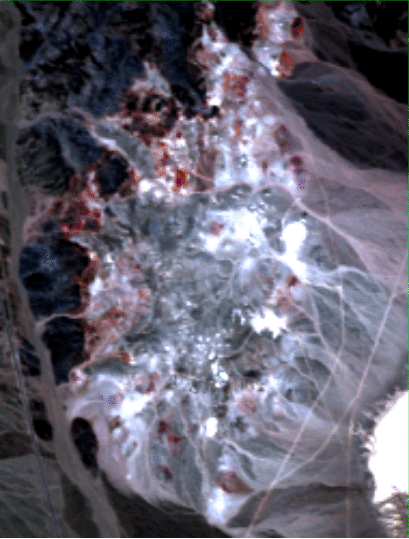}\\
    \centering \small Cuprite HI
    \end{minipage}

    \vspace{0.1cm}
    \caption{Visual representation of the real HIs used in the experiments.} %
    \label{fig:realData_RGB_imgs}
\end{figure}

\begin{figure*}
    \centering
    \includegraphics[height=0.32\pdfpageheight]{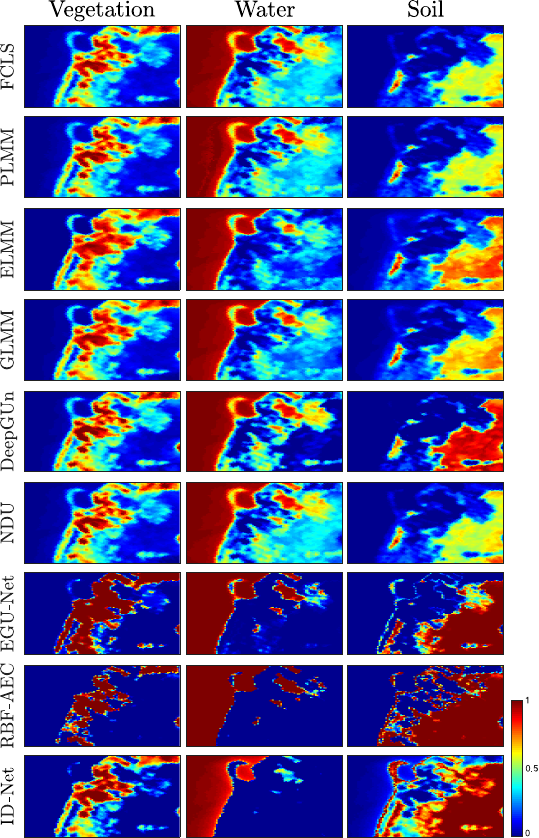}
    \hspace{0.01\linewidth}
    \includegraphics[height=0.32\pdfpageheight]{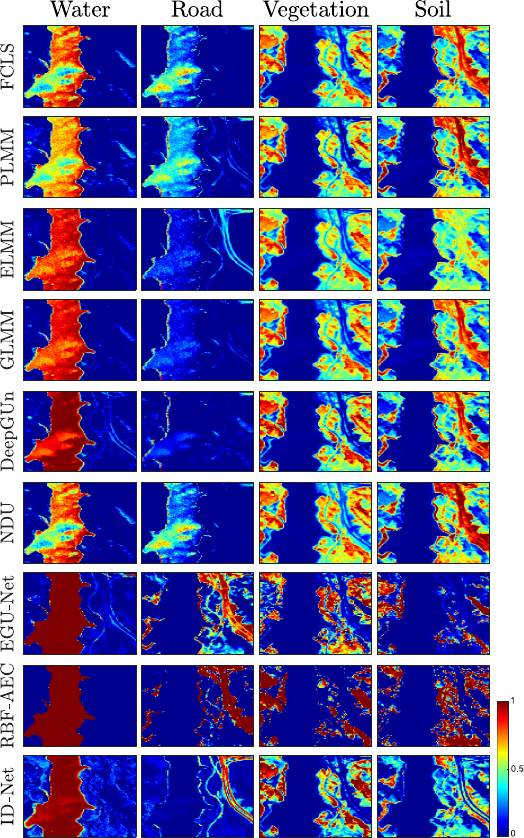}
    \hspace{0.01\linewidth}
    \includegraphics[height=0.32\pdfpageheight]{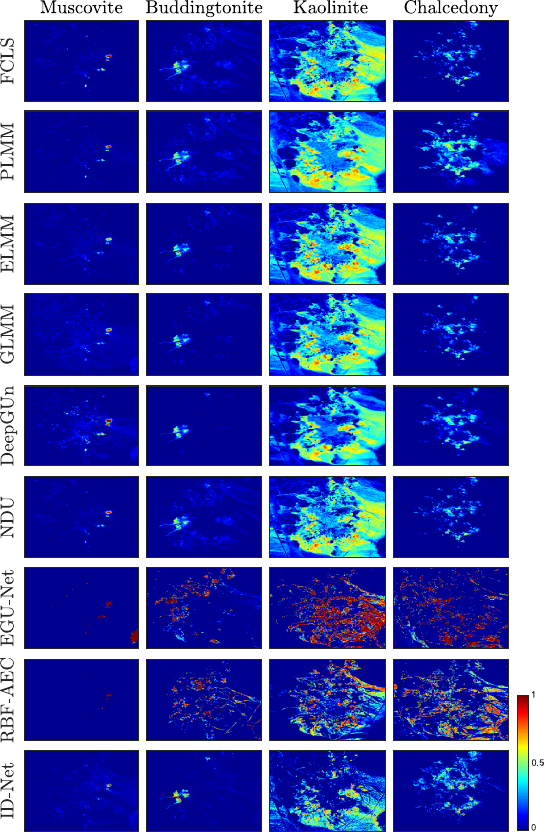}
    \vspace{-0.2cm}
    \caption{Reconstructed abundance maps for the experiments with real data for the Samson (left), Jasper Ridge (middle) and Cuprite (right) HIs.}
    \label{fig:ex_real_abunda}
\end{figure*}

\subsection{Real data}

We also evaluate the performance of the methods on the Samson, on the Jasper Ridge, and on the Cuprite HIs. These images were acquired by the AVIRIS instrument, which samples the spectra on 224 spectral bands. Bands corresponding to water absorption regions or with a low SNR were removed from all datasets, which resulted in 156 bands for the Samson HI, 198 bands for the Jasper Ridge HI, and 188 bands for the Cuprite HI. Previous works have shown that the Samson and Jasper Ridge HIs contain three and four EMs, respectively~\cite{borsoi2018superpixels1_sparseU,imbiriba2018ULTRA_V}, while fourteen EMs are used for the Cuprite HI~\cite{Nascimento2005,bhatt2014dataDrivenUnmixing}.
These datasets are relevant since they represent nonidealities such as EM variability, which can appear due to non-trivial variations in illumination from irregular terrain topography and due to the intrinsic variations of the materials such as soil, vegetation and road, and nonlinear mixtures, which are often present in mixtures containing vegetation or water, or in mixtures of minerals. A representation of the real datasets can be seen in Figure~\ref{fig:realData_RGB_imgs}.

\begin{figure}
    \centering
    \includegraphics[width=0.75\linewidth]{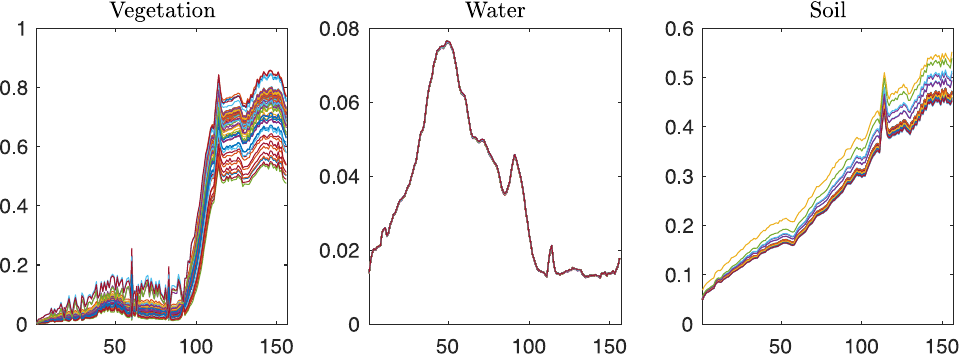}
    \\
    \vspace{0.3cm}
    \includegraphics[width=0.95\linewidth]{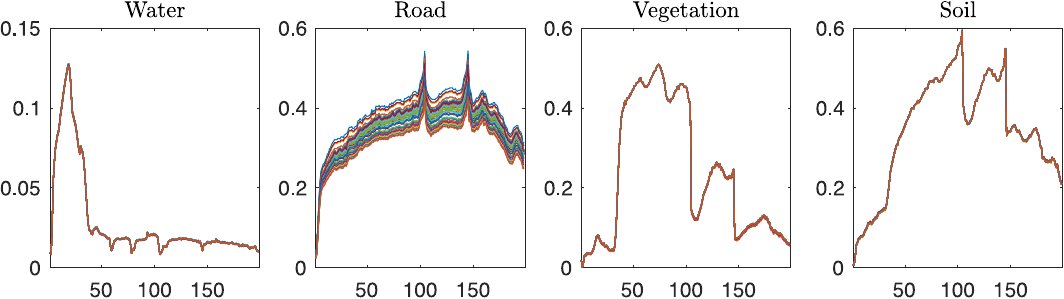}
    \\
    \vspace{0.3cm}
    \includegraphics[width=\linewidth]{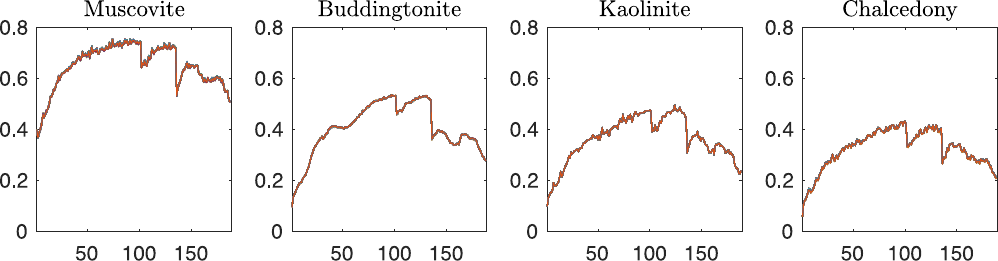}
    \vspace{-0.6cm}
    \caption{Samples of the EMs estimated by ID-Net for the Samson (top), Jasper Ridge (middle), and Cuprite (bottom) HIs.}
    \label{fig:est_EMs_real}
\end{figure}

\begin{figure}
    \centering
    \includegraphics[height=0.245\linewidth]{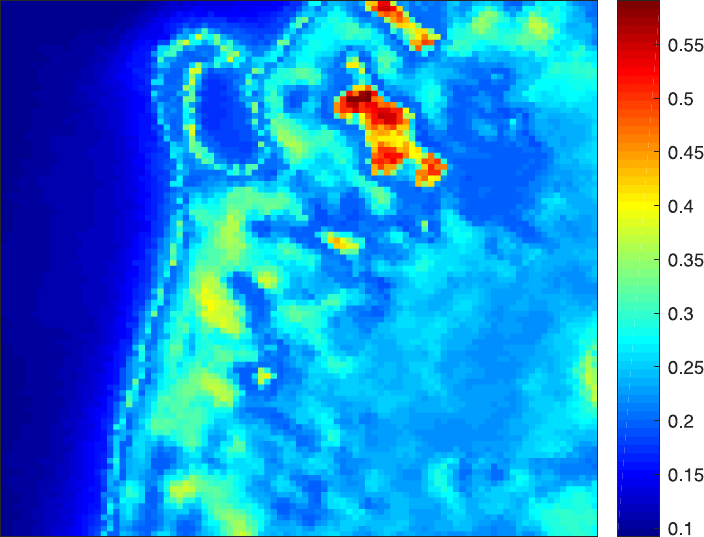}
    \includegraphics[height=0.245\linewidth]{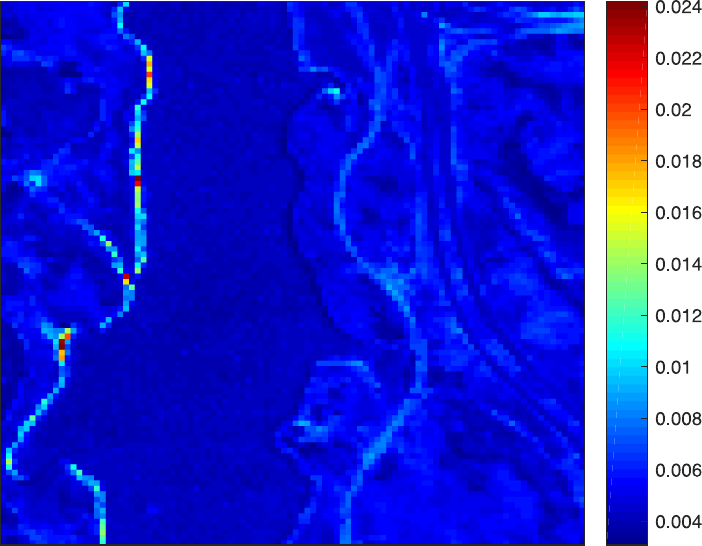}
    \includegraphics[height=0.245\linewidth]{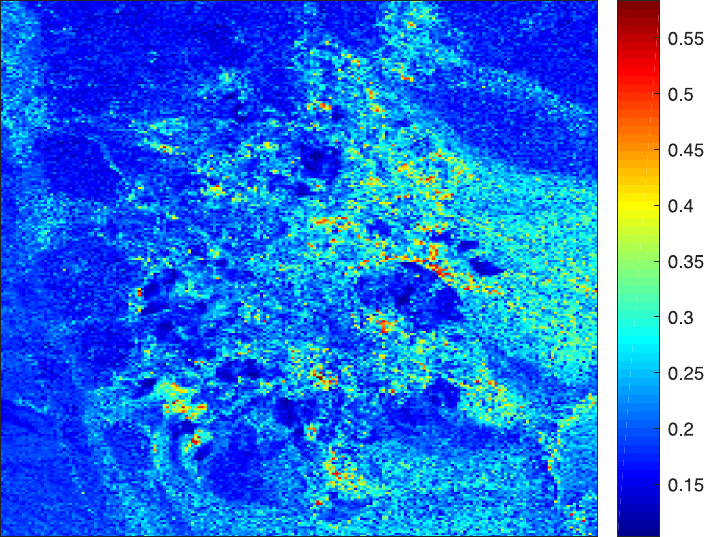}

    \vspace{-0.15cm}
    \caption{Nonlinearity degree $\eta_d$ of the HU process in ID-Net for the Samson (left), Jasper Ridge (middle) and Cuprite (right) HIs.}
    \label{fig:nonlinearityDeg}
\end{figure}

The abundances maps reconstructed by the different algorithms and datasets are presented in Figure~\ref{fig:ex_real_abunda}. 
For the Samson and Jasper Ridge images, it can be noticed that the methods based on deep-learning (particularly, EGU-Net, RBF-AEC and ID-Net) generally provided a better separation between the different materials in the scene when compared to the remaining algorithms. This can be observed by the smaller confusion in the results by those methods between Water and Soil materials for the Samson HI, and between Road and Soil EMs for the Jasper Ridge HI.
For the Cuprite HI (for which only the Muscovite, Buddingtonite, Kaolinite and Chalcedony EMs are shown), a similar behavior is observed for the learning-based methods. The abundances obtained by most of the algorithms generally agree with previous studies on this HI~\cite{Nascimento2005,bhatt2014dataDrivenUnmixing}, however, for some EMs (the Buddingtonite and Chalcedony) the abundances retrieved by EGU-Net and RBF-AEC show a different spatial distribution compared to the remaining algorithms. However, pixel-by-pixel comparisons on this scene are hard due to the lack of ground truth~\cite{kruse2003comparisonAvirisHypherionCuprite}.
When compared to the competing methods, ID-Net was able to obtain a better abundance reconstruction in regions with more heavily mixed EMs. This can be seen more clearly in the Vegetation and Soil materials in the Samson HI, for which the results of EGU-Net and RBF-AEC identify most pixels as being completely pure. 
Samples of the EMs estimated by ID-Net are shown in Figure~\ref{fig:est_EMs_real}. It can be seen that the amount of variability was different for each material and image, being higher in the Samson HI, or for vegetation, compared to the Cuprite HI, or water.
It is also instructive to analyze the nonlinearity degree (i.e., the contribution of the nonlinear part of the NN) in the estimated abundance coefficients in~\eqref{eq:abundances_posterior_twoStream} for each pixel, defined here as $\eta_d=\|\breve{\bgamma}_{\phi}^{a,{\rm nlin}}(\by,\bM)\|\big/\big(\|\breve{\bgamma}_{\phi}^{a,{\rm lin}}(\by,\bM)\| + \|\breve{\bgamma}_{\phi}^{a,{\rm nlin}}(\by,\bM)\|\big)$, which is shown in Figure~\ref{fig:nonlinearityDeg}. It can be seen that the overall level of nonlinearity degree varies across the HI. As expected, $\eta_d$ is generally higher in areas with more mixed pixels, and lower in homogeneous areas with predominantly pure pixels. This is particularly seen in areas containing mixtures of vegetation and water in the Samson HI. However, we emphasize that nonlinearity in~\eqref{eq:abundances_posterior_twoStream} and EM variability jointly influence the HU results, and separating the effect of each one is not trivial.
The quantitative results of the algorithms are shown in Table~\ref{tab:results_realData}. It can be seen that the GLMM and NDU methods obtained the smallest reconstruction errors ($\text{NRMSE}_{\bY}$). This occurs since these methods have a high amount of degrees of freedom and can represent HI pixels very closely. 
However, we note that achieving low values of $\text{NRMSE}_{\bY}$ do not imply good abundance estimation performance.
The execution times of ID-Net were generally higher but still comparable to those by DeepGUn and NDU (1.33 and 1.26 times on average, respectively). This is a limitation of the proposed method, and in future work we will investigate more efficient inference techniques to address it.

The algorithms based on deep learning led to distinct results due to their different frameworks.
RBF-AEC and EGU-Net, which are based on AEC NN architectures, obtained abundances that were generally sparse, with many pure components. 
DeepGUn, which is based on a matrix factorization framework with generative EM models, led to abundances comparatively closer to model-based methods such as the GLMM. ID-Net, which leverages a statistical model of the unmixing processes, gave results with pure components in homogeneous regions and mixed pixels in the regions of transitions between materials.
Thus, while the differences between the results of these methods are not straightforward to explain, we can observe how they change as we transition from model-based towards fully neural network-based algorithms.

\section{Conclusions}
\label{sec:conclusions}

This paper proposed a hyperspectral unmixing method considering both nonlinearity and EM variability based on a deep disentangled variational inference framework. Both the model parameters and a tractable approximation of the posterior distribution were learned end-to-end by maximizing a lower bound to the likelihoods of supervised and unsupervised data. A self-supervised learning strategy was considered to leverage the benefits of semi-supervised learning algorithms while using only data extracted from the observed image. The abundances and EMs were disentangled through the use of appropriate independence assumptions on the variational posterior. Moreover, an interpretable model was designed by using unrolled optimization-based and two-stream (linear/nonlinear) NN architectures, allowing for an adjustable amount of nonlinearity in the model. Experimental results on both synthetic and real datasets showed that the proposed method can achieve state-of-the-art performance with execution times comparable to that of algorithms such as NDU and DeepGUn.
Several interesting directions for future work can be considered, such as developing more efficient inference strategies that lead to algorithms with smaller computational cost; designing self-supervised learning approaches to extract the training data from the HI which do not depend on the presence of multiple pure pixels; and exploiting the statistical dependence among abundances of spatially adjacent pixels.

\bibliographystyle{IEEEtran}
\bibliography{references_vaeU,references_varRev2}

\end{document}